\documentclass[12pt,a4,portrait]{article}

\usepackage{latexsym}
\usepackage{hyperref}
\usepackage{mathtools}
\usepackage{multirow}
\usepackage{mathrsfs}
\usepackage{cite}
\usepackage[T1]{fontenc}
\usepackage{tocbibind}

\begin{document}

\title{Transformation properties and general relativity regime in scalar-tensor theories}
\author{Laur J\"arv\thanks{laur.jarv@ut.ee}, Piret Kuusk\thanks{piret.kuusk@ut.ee}, Margus Saal\thanks{margus.saal@ut.ee} and Ott Vilson\thanks{ovilson@ut.ee} \\
{\normalsize Institute of Physics, University of Tartu, Ravila 14c, Tartu 50411, Estonia}} 

%\address{Institute of Physics, University of Tartu, Ravila 14c, Tartu 50411, Estonia}

%\eads{\mailto{laur.jarv@ut.ee}, \mailto{piret.kuusk@ut.ee}, \mailto{margus.saal@ut.ee} and \mailto{ovilson@ut.ee}}

\date{}
\maketitle

\begin{abstract}
We consider first generation scalar-tensor theories of gravitation in a completely generic form, keeping the transformation functions of the local rescaling of the metric and the scalar field redefinition explicitly distinct from the coupling functions in the action.
It is well known that in the Jordan frame Brans-Dicke type parametrization the diverging kinetic coupling function $\omega \rightarrow \infty$ can lead to the general relativity regime, however then the transformation functions to other parametrizations typically become singular, possibly spoiling the correspondence between different parametrizations.
We give a detailed analysis of the transformation properties of the field equations
with arbitrary metric and also in the Friedmann cosmology, and provide sufficient conditions under which the correspondence between different parametrizations is retained, even if the transformation is singular.
It is interesting to witness the invariance of the notion of the general relativity regime and the correspondence of the perturbed cosmological equations as well as their solutions in different parametrizations, despite the fact that in some cases the perturbed equation turns out to be linear in one parametrization and nonlinear in some other.
\vspace{0.2cm}\\
Keywords: scalar-tensor gravity, general relativity limit, Friedmann cosmology, transformation properties\\
PACS numbers: 04.50.Kd, 98.80.Jk

\end{abstract}
%\pacs{04.50.Kd, 98.80.Jk}

%\noindent{\it Keywords}: scalar-tensor gravity, general relativity limit, Friedmann cosmology, transformation properties

%\submitto{\CQG}

\section{Introduction}

The history of the scalar-tensor theory of gravitation (STG) \cite{stg:books,DamourEF} as an extension to Einstein's general relativity (GR) in principle started with the works by Kaluza and Klein. Complementary ideas were pursued by Jordan and Fierz \cite{Jordan}, developed by Brans and Dicke \cite{BransDicke} and further generalized by Bergmann and Wagoner \cite{bergmann,wagoner}. Nowadays aforementioned can be called first generation scalar-tensor theories. The Horndeski theory \cite{Horndeski} which also allows derivative couplings and possesses equations of motion with up to second order derivatives of the metric and scalar field, may be considered to be the second generation. Healthy, ghost free theories going beyond Horndeski can be referred to as the third generation \cite{disformal}.

Soon after his joint work with Brans \cite{BransDicke}, Dicke published another paper \cite{Dicke} where he recalled the local Weyl rescaling of the metric tensor, interpreted it as a transformation of the units and claimed that physics must be invariant under this transformation \cite{Faraoni:2006,Flanagan}. From that viewpoint STG is a natural extension of GR because rewriting the Einstein-Hilbert action in terms of a Weyl rescaled metric tensor introduces an action functional having a structure that resembles the one used for STG \cite{TsamisWoodard,deruelle:sasaki}. Namely, on the level of the rewritten action functional the scalar field entering via the Weyl rescaling is coupled to curvature, to matter etc. Of course in that case the functions describing the coupling of the scalar field to curvature etc.~are related to each other in a specific way which implies that the scalar field equation of motion is an identity $0 \equiv 0$, and hence the scalar field is not a physical degree of freedom. Nevertheless if one considers an analogous action functional but without the relations between the coupling functions then the resulting theory is STG, congruent with Weyl integrable geometry \cite{romero}.

Already Jordan \cite{Jordan} pointed out that for scalar-tensor theories with a constant kinetic coupling parameter $\omega$ the equations of motion reduce to those of GR if $\omega = \infty$. In the framework of the parametrized post-Newtonian approximation it was shown that for the theory with a dynamical $\omega \equiv \omega(\Psi)$ \cite{bergmann,wagoner} conditions for the theory to comply with GR is again $\omega(\Psi) \to \infty$ as well as $\frac{\omega^\prime(\Psi)}{\omega(\Psi)^3} \to 0$ \cite{Nordtvedt,Hohmann:2013rba}. In the context of the Friedmann cosmology Damour and Nordtvedt \cite{DamourNordtvedt:1993a,DamourNordtvedt:1993b} showed that for a wide family of theories the limit $\omega(\Psi) \to \infty$ is an attractor. To be more precise there exists a mechanism ending the scalar field evolution at some constant value thereby rendering the remaining dynamical degrees of freedom identical to those of GR. In the current paper we shall use the term `GR regime' to refer to such situation. Due to these results a dynamical approach to the GR regime has been studied by number of authors, e.g.~\cite{Barrow:Parsons,Barrow:Shaw,Minazzoli:Hees,Serna:2002fj,Modak:etal,MaedaFujii,Billyard,Bhadra,MimosoWands,Garcia:Wands}. 

Damour and Nordtvedt noted that the points in the field space where $\omega(\Psi) = \infty$ enter the theory as mathematically singular boundary points \cite{DamourNordtvedt:1993b}. They used the local Weyl rescaling of the metric tensor and redefined (reparametrized) the scalar field $\varphi = \varphi(\Psi)$ in order to rewrite the theory in the so called Einstein frame where all functions are regular. However, the singularity in $\omega(\Psi) \to \infty$ (in the so called Jordan frame) is then absorbed by the scalar field redefinition hence rendering the transformation to be singular instead, i.e.~$\frac{\mathrm{d}\varphi}{\mathrm{d}\Psi}\to \infty$. Therefore, it is not so obvious that these transformations can be trusted at all and one must take extra caution when applying the transformation in the vicinity of the GR regime \cite{JKS2007}. 

Note that in the literature when the equivalence of the parametrizations is discussed then the transformation functions are often assumed to be regular \cite{Faraoni:2006,Flanagan} which in principle is easily achievable when a suitable choice of coupling functions in the Jordan frame is considered. However in our recent paper \cite{JKSV:2} we showed that the scalar field $\Psi$ in the Jordan frame is equivalent to the invariant notion of the nonminimal coupling while the Einstein frame scalar field $\varphi$ is equivalent to the invariant notion of the scalar field space volume. Therefore $\frac{\mathrm{d}\Psi}{\mathrm{d}\varphi}=0$ in the GR regime is not due to an unfortunate choice of coupling functions but it is a crucial part of the notion of the GR regime, stating via invariants that the nonminimal coupling vanishes. Hence we conclude that the singular scalar field redefinition is physically meaningful and deserves a closer look. 

In the current paper we intend to clarify the question whether or not such a singular transformation is permitted by first studying the transformation properties of the action, the equations of motion and the Friedmann cosmology. Afterwards we focus upon the transformations in the neighbourhood of the GR regime which corresponds to a critical point of the scalar field equation of motion. We argue that the conditions for critical points in general as well as in the Friedmann cosmology are preserved under the scalar field redefinition even if the latter is singular. Most importantly we show in detail that the perturbed equation, approximating the dynamics in the vicinity of the GR regime, transforms well despite the fact that in the case of the singular scalar field redefinition a nonlinear perturbed equation gets transformed into a linear one. The transformation of the solutions also shows an analogous interesting correspondence. To give a completely generic treatment of the transformations between all possible parametrizations we adopt the notation introduced by Flanagan \cite{Flanagan}. The paper accords with the spirit of recent works \cite{Chiba:Yamaguchi,Domenech:Sasaki} etc.~where the correspondence between Jordan and Einstein frames is discussed in explicit details.

The outline of the paper is the following. In Section~\ref{general.theory} we write down the action functional, derive the equations of motion and plug in the Friedmann-Lema\^itre-Robertson-Walker (FLRW) line element in order to obtain the general Friedmann cosmology in the context of STG. In Section~\ref{general.relativity.regime} we introduce the notion of the general relativity regime by examining the necessary conditions for maintaining the constancy of the scalar field once it has been obtained. Section~\ref{dynamical.system} completes the line of thought of \cite{JKS2007,JKS2008,JKS:2007:etc,JKS2010,JKS:2010:time} by considering a dynamical approach to the general relativity regime in the context of the potential dominated epoch of the Friedmann cosmology. In the current paper the latter serves as an example for showing the equivalence of different parametrizations on the level of the perturbed equations. It turns out to be nontrivial and we have included a lot of calculational details in order to keep the treatment as traceable as possible.

From the structural point of view the paper is divided into three sections each of which is split into two halves. In the first halves of the sections a relatively complete theory in an arbitrary parametrization starting by the action functional and ending with the solutions in the context of the Friedmann cosmology is given. The second halves follow the first halves by providing the corresponding transformation properties under the local Weyl rescaling of the metric tensor and under the scalar field redefinition. Therefore subsections numbered as $i.1.j$ contain the theory and $i.2.k$ discuss the transformation properties of the quantities introduced in $i.1.j$. 

%%%%%%%%%%%%%%%%%%%%%%%%%%%%%%%%%%%%%%%%%%%%%%%%%%%%%%%%%%%%%%%%%%%%%%%%%%%%%%%%%

\section{General theory}\label{general.theory}

In this section we write down an action functional and derive the equations of motion. Also the general Friedmann cosmology is discussed.

%%%%%%%%%%%%%%%%%%%%%%%%%%%%%%%%%%%%%%%%%%%%%%%%%%%%%%%%%%%%%%%%%%%%%%%%%%%%%%%%%

\subsection{Theory: part I}

%%%%%%%%%%%%%%%%%%%%%%%%%%%%%%%%%%%%%%%%%%%%%%%%%%%%%%%%%%%%%%%%%%%%%%

\subsubsection{Action functional}

Let us consider a family of theories of gravitation by postulating an action functional \cite{Shapiro:Takata,Flanagan}
\begin{equation}
\label{fl.moju}
S = \frac{1}{2\kappa^2}\int_{V_4}\mathrm{d}^4x\sqrt{-g}\left\lbrace {\mathcal A}(\Phi)R-
{\mathcal B}(\Phi)g^{\mu\nu}\nabla_\mu\Phi \nabla_\nu\Phi - 2\ell^{-2}{\mathcal V}(\Phi)\right\rbrace 
+ S_\mathrm{m}\left[\mathrm{e}^{2\alpha(\Phi)}g_{\mu\nu},\chi^{A}\right] \,.
\end{equation}
There are two unspecified constants: $\kappa^2$ yields the dimension for the gravitational ``constant'' and $\ell>0$ has the dimension of length. We make use of the convention $c \equiv 1$ and have suitably chosen constants $\kappa^2$ and $\ell^{-2}$ in order to consider the scalar field $\Phi$ and the four arbitrary functions $\left\lbrace \mathcal{A}(\Phi),\,\mathcal{B}(\Phi),\,\mathcal{V}(\Phi),\,\alpha(\Phi) \right\rbrace$ of it to be dimensionless, regardless whether in addition either $\kappa^2 \equiv 1$ or $\hbar \equiv 1$ is imposed.

Note that in the general case the action functional $S_\mathrm{m}$ for the matter fields $\chi^{A}$, where different components are labelled by the superscript $A$, functionally depends on the metric tensor $\hat{g}_{\mu\nu} = \mathrm{e}^{2\alpha(\Phi)}g_{\mu\nu}$. Nevertheless the coupling of the matter fields to the geometry described by $g_{\mu\nu}$ is universal and therefore one of the basic principles underlying the general relativity is fulfilled.

In order to consider a concrete theory one must specify each of the four arbitrary functions $\left\lbrace \mathcal{A}(\Phi),\, \mathcal{B}(\Phi),\, \mathcal{V}(\Phi),\, \alpha(\Phi) \right\rbrace$. However in the literature mostly such action functionals have been considered where the functional form of two out of the four arbitrary functions has been specified because in that case the calculations are easier, while the corresponding action functional has retained its generality up to some details \cite{Faraoni:no:progress,JKSV:2}. In the current paper we shall use `parametrization' to refer to these setups and 
hereby recall two most well known ones:

\begin{itemize}
	\item The Jordan frame action in the Brans-Dicke-Bergmann-Wagoner parametrization (JF BDBW) 
	\cite{BransDicke,bergmann,wagoner} for the 
	scalar field denoted as $\Psi$ is obtained as follows:
	\begin{equation}
	\label{Jordan.frame}
	{\mathcal A} = \Psi\,,\, {\mathcal B} = \frac{\omega(\Psi)}{\Psi} \,,\, {\mathcal V} = \mathcal{V}_\mathrm{J}(\Psi)\,,\, \alpha = 0 \,.
	\end{equation}
	
	\item  The Einstein frame action in canonical parametrization (EF can) \cite{Dicke,bergmann,wagoner} 
	for the scalar field denoted as $\varphi$ is obtained as follows:
	\begin{equation}
	\label{Einstein.frame}
	 {\mathcal A} = 1 \,,\, {\mathcal B} = 2\,,\, {\mathcal V} = \mathcal{V}_\mathrm{E}(\varphi)\,,\, \alpha = \alpha_\mathrm{E}(\varphi) \,.
	\end{equation}

\end{itemize}
Here and in the following we shall drop the arguments of the arbitrary functions $\left\lbrace \mathcal{A}(\Phi),\,\mathcal{B}(\Phi),\right.$ $\left. \mathcal{V}(\Phi),\,\alpha(\Phi) \right\rbrace$ unless confusion might arise. We also adopt a convention where prime means derivative w.r.t.~the scalar field, e.g.
\begin{equation}
\label{convention}
\mathcal{A}^\prime \equiv \frac{\mathrm{d}\mathcal{A}(\Phi)}{\mathrm{d}\Phi} \,,\, \mathcal{B}^\prime \equiv \frac{\mathrm{d}\mathcal{B}(\Phi)}{\mathrm{d}\Phi} \,,\,\text{etc.}
\end{equation}
In the current paper we shall use the so-called mostly plus signature for the metric tensor $g_{\mu\nu}$ and always assume the affine connection to be the Levi-Civita one. The other unspecified conventions are as e.g.~in the textbook by Carroll \cite{carroll}. 

%%%%%%%%%%%%%%%%%%%%%%%%%%%%%%%%%%%%%%%%%%%%%%%%%%%%%%%%%%%%%%%%%%%%%%

\subsubsection{Equations of motion}

Varying the action \eqref{fl.moju} while considering $g^{\mu\nu}$, $\Phi$ and $\chi^{A}$ to be the dynamical fields reads
\begin{align}
\nonumber
\delta S &= \frac{1}{2\kappa^2}\int_{V_4} \mathrm{d}^4x  \sqrt{-g} \left\lbrace E^{(g)}_{\mu\nu} \delta g^{\mu\nu} + E^{(\Phi)} \delta\Phi + 2\kappa^2 \mathrm{e}^{4\alpha} E^{(\chi)}_{A} \delta\chi^{A} \right\rbrace  \\
\label{varied.action}
&\quad+ \frac{1}{2\kappa^2}\int_{V_4} \mathrm{d}^4x\partial_\sigma \left( \sqrt{-g} \left[ \mathscr{B}_{(g)}^{\,\sigma} + \mathscr{B}_{(\Phi)}^{\,\sigma}  + 2 \kappa^2 \mathrm{e}^{4\alpha} \mathscr{B}_{(\chi)}^{\,\sigma}  \right] \right)
\end{align}
where
\begin{align}
\label{boundary.term.metric}
\sqrt{-g}\, \mathscr{B}_{(g)}^{\,\sigma} &= \sqrt{-g} \left\lbrace \mathcal{A}g_{\mu\nu} g^{\sigma\lambda}  \nabla_{\lambda} \delta g^{\mu\nu} - \mathcal{A} \nabla_\mu \delta g^{\sigma\mu} - g^{\sigma\lambda} \left( \nabla_\lambda \mathcal{A} \right) g_{\mu\nu} \,\delta g^{\mu\nu} + \left( \nabla_\mu \mathcal{A} \right) \delta g^{\mu\sigma} \right\rbrace \,,\\
\label{boundary.term.scalar.field}
\sqrt{-g}\, \mathscr{B}_{(\Phi)}^{\,\sigma} &= -\sqrt{-g}\, 2\mathcal{B}g^{\sigma\mu}\left( \nabla_\mu \Phi \right) \delta \Phi
\end{align}
and $\sqrt{-g}\,\mathrm{e}^{4\alpha} \mathscr{B}_{(\chi)}^{\,\sigma}$ are eventually the boundary terms arising from varying w.r.t.~the metric tensor $g^{\mu\nu}$, w.r.t.~the scalar field $\Phi$ and w.r.t.~the matter fields respectively. The boundary terms have been written out explicitly for the sake of completeness, although they do not give a contribution to the equations of motion. Therefore by making use of the minimal action principle $\delta S = 0$ we obtain the equations of motion as follows:
\begin{align}
\nonumber
E^{(g)}_{\mu\nu} \equiv& \mathcal{A}\left(R_{\mu\nu} - \frac{1}{2}g_{\mu\nu}R\right) + \left( \frac{1}{2}\mathcal{B}+\mathcal{A}^{\prime\prime}\right) g_{\mu\nu} g^{\rho\sigma}\nabla_\rho\Phi\nabla_\sigma\Phi - \left( \mathcal{B} + \mathcal{A}^{\prime\prime} \right)\nabla_\mu\Phi\nabla_\nu\Phi \\
\label{tensor.equation}
&+ \mathcal{A}^\prime\left( g_{\mu\nu}\Box\Phi - \nabla_\mu\nabla_\nu\Phi\right) + \ell^{-2} g_{\mu\nu}\mathcal{V} - \kappa^2T_{\mu\nu} = 0\,, \\
\label{scalar.field.equation.containing.R}
E^{(\Phi)} \equiv& R\mathcal{A}^\prime + \mathcal{B}^\prime g^{\mu\nu}\nabla_\mu\Phi\nabla_\nu\Phi + 2\mathcal{B}\Box\Phi - 2\ell^{-2}\mathcal{V}^\prime + 2\kappa^2\alpha^\prime T = 0  \,, \\
\label{matter.eom}
E^{(\chi)}_A \equiv& E^{(\chi)}_A\left[ \mathrm{e}^{2\alpha}g_{\mu\nu} , \chi^{C} \right] = 0 \,.
\end{align}
Here
\begin{equation}
\label{energy.momentum.tensor}
T_{\mu\nu} \equiv -\frac{2}{\sqrt{-g}}\frac{\delta S_\mathrm{m}}{\delta g^{\mu\nu}}
\end{equation}
is the matter energy-momentum tensor, $T \equiv g^{\mu\nu}T_{\mu\nu}$ is its contraction and $\Box \equiv g^{\mu\nu}\nabla_\mu\nabla_\nu$. In the current paper we are not directly interested in the equations of motion for the matter fields $\chi^{A}$, i.e.~we do not specify neither \eqref{matter.eom} nor the corresponding boundary terms. However including them provides us a complete picture at least on the schematic level and allows us to stress an important point. Namely, the matter fields $\chi^A$ ``feel'' the geometry determined by $\hat{g}_{\mu\nu} \equiv \mathrm{e}^{2\alpha}g_{\mu\nu}$. Therefore freely falling material objects follow the corresponding geodesics. Hence if one intends to measure the geometry determined by $g_{\mu\nu}$ using reference objects built out of the matter fields then, in the spirit of Dicke \cite{Dicke}, correction factors must be applied.

In the literature usually the contraction of \eqref{tensor.equation}, i.e.
\begin{equation}
\label{contraction.of.tensor.equation}
g^{\mu\nu} E^{(g)}_{\mu\nu} \equiv -\mathcal{A}R + \mathcal{B}g^{\mu\nu} \nabla_\mu \Phi \nabla_\nu \Phi + 3 \mathcal{A}^{\prime\prime}g^{\mu\nu} \nabla_\mu \Phi \nabla_\nu \Phi + 3\mathcal{A}^\prime \Box \Phi + 4 \ell^{-2} \mathcal{V} - \kappa^2 T = 0
\end{equation}
is used to eliminate the Ricci scalar $R$ from \eqref{scalar.field.equation.containing.R} in order to obtain an equation of motion for the scalar field $\Phi$ that does not contain the second derivatives of the metric tensor $g_{\mu\nu}$ and therefore purely describes the propagation of the scalar field. The result reads 
\begin{equation}
\label{scalar.field.equation.without.R}
\frac{2\mathcal{A}\mathcal{B} \negmedspace + \negmedspace 3\left( \mathcal{A}^\prime\right)^2}{\mathcal{A}}\Box\Phi + \frac{\left( 2\mathcal{A}\mathcal{B} \negmedspace + \negmedspace 3\left( \mathcal{A}^\prime \right)^2 \right)^\prime}{2\mathcal{A}} g^{\mu\nu} \nabla_\mu\Phi \nabla_\nu\Phi - \frac{2\left( \mathcal{A}\mathcal{V}^\prime \negmedspace - \negmedspace 2 \mathcal{V} \mathcal{A}^\prime \right)}{\ell^2\mathcal{A}} + \frac{\kappa^2\left( 2\mathcal{A}\alpha^\prime \negmedspace - \negmedspace \mathcal{A}^\prime \right)}{\mathcal{A}}T = 0\,.
\end{equation}
This procedure is also known as `debraiding', see e.g.~a recent paper by Bettoni {\it et al} \cite{bettoni} for comments and further references. Note that due to $\nabla_\mu \nabla_\nu \Phi$ in \eqref{tensor.equation} it is not possible to make an analogous substitution to obtain an equation that would describe solely the evolution of the metric tensor. In some sense this is the underlying motivation for the Einstein frame canonical parametrization \eqref{Einstein.frame}. Last but not least combining \eqref{scalar.field.equation.containing.R} and the covariant divergence of the tensor equation \eqref{tensor.equation} leads us to
\begin{equation}
\label{EI.jaavus}
E^{(c)}_{\nu} \equiv \nabla^\mu E^{(g)}_{\mu\nu} + \frac{1}{2} E^{(\Phi)} \nabla_\nu\Phi = - \kappa^2\nabla^\mu T_{\mu\nu} + \kappa^2\alpha^\prime T \nabla_\nu\Phi = 0 \,
\end{equation}
which is the well known continuity equation.

%%%%%%%%%%%%%%%%%%%%%%%%%%%%%%%%%%%%%%%%%%%%%%%%%%%%%%%%%%%%%%%%%%%%%%

\subsubsection{Friedmann cosmology}\label{Friedmann.cosmology}

Let us consider the FLRW line element in spherical coordinates
\begin{equation}
\label{FLRW}
\mathrm{d}s^2 = -\mathrm{d}t^2 + \left(a(t)\right)^2\left( \frac{\mathrm{d}r^2}{1 - kr^2} + r^2 \mathrm{d} \Omega^2 \right)
\end{equation}
defined in an arbitrary parametrization. Here $t$ and $a(t)$ are respectively the cosmological time and the scale factor connected to the chosen parametrization. The constant $k$ takes values $-1$, $0$ and $+1$ determining the spatial geometry to be hyperbolic, flat or spherical respectively. The dependence on the two angles is gathered into $\mathrm{d}\Omega^2$. Due to the homogeneity and isotropy assumption underlying the Friedmann cosmology the scalar field can only depend on the cosmological time $\Phi\equiv\Phi(t)$. The equations of motion \eqref{tensor.equation}, \eqref{scalar.field.equation.without.R} and \eqref{EI.jaavus} in the case of FLRW metric read
\begin{align}
\label{First.Friedmann.general.equation} H^2 &= -\frac{\mathcal{A}^\prime}{ \mathcal{A} } H \dot{\Phi} + \frac{ \mathcal{B} }{ 6\mathcal{A} } \dot{\Phi}^2 + \frac{1}{3\ell^2 \mathcal{A}}\mathcal{V} + \frac{\kappa^2}{3\mathcal{A}}\rho - \frac{k}{a^2} \,, \\
\label{Second.Friedmann.general.equation} \negmedspace 2\dot{H} \negmedspace + \negmedspace 3H^2 &= -2\frac{\mathcal{A}^\prime}{\mathcal{A}} H \dot{\Phi} - \left( \frac{\mathcal{B}}{2\mathcal{A}} + \frac{\mathcal{A}^{\prime\prime}}{\mathcal{A}}\right)  \dot{\Phi}^2 - \frac{\mathcal{A}^\prime}{\mathcal{A}}\ddot{\Phi} + \frac{1}{\ell^2\mathcal{A}}\mathcal{V}  - \frac{\kappa^2}{\mathcal{A}}p - \frac{k}{a^2} \,, \\
\label{Scalar.field.equation.in.FLRW.cosmology}
\ddot{\Phi} &= -3 H\dot{\Phi} \negmedspace - \negmedspace \frac{1}{2}\frac{\left(2 \mathcal{A} \mathcal{B} \negmedspace + \negmedspace 3 (\mathcal{A}^\prime)^2 \right)^\prime}{\left(2 \mathcal{A} \mathcal{B} \negmedspace + \negmedspace 3 (\mathcal{A}^\prime)^2 \right)}\dot{\Phi}^2 \negmedspace - \negmedspace 2\ell^{-2}\frac{ \mathcal{A} \mathcal{V}^\prime \negmedspace -\negmedspace 2 \mathcal{V} \mathcal{A}^\prime }{2 \mathcal{A} \mathcal{B} \negmedspace + \negmedspace 3 (\mathcal{A}^\prime)^2} \negmedspace - \negmedspace \kappa^2\frac{\left( 2\mathcal{A}\alpha^\prime \negmedspace - \negmedspace \mathcal{A}^\prime \right)}{2 \mathcal{A} \mathcal{B} \negmedspace + \negmedspace 3 (\mathcal{A}^\prime)^2}\left( \rho - 3p \right)\,, \\
\label{cosmology.continuity.equation}
\phantom{\frac{A}{A}}\dot{\rho} &=-3H\left(\rho + p\right) + \alpha^\prime \left(\rho - 3p \right)\dot{\Phi} \,,
\end{align}
where dot means derivative with respect to the cosmological time $t$ and $H \equiv \frac{\dot{a}}{a}$ is the Hubble parameter. We have assumed the matter to be a perfect fluid with the energy density $\rho$ and pressure $p$.

%%%%%%%%%%%%%%%%%%%%%%%%%%%%%%%%%%%%%%%%%%%%%%%%%%%%%%%%%%%%%%%%%%%%%%%%%%%%%%%%%%%%%%%%%%%%%%%%%%%%%%%%%%%%%%%%%%

\subsection{Transformations: part I} 

%%%%%%%%%%%%%%%%%%%%%%%%%%%%%%%%%%%%%%%%%%%%%%%%%%%%%%%%%%%%%%%%%%%%%%

\subsubsection{Transformation of the action functional}

It is well known that the action functional \eqref{fl.moju} preserves its structure up to a boundary term under the transformations that contain two functional degrees of freedom
\begin{align}
\label{conformal.transformation}
g_{\mu\nu} &= \mathrm{e}^{2\bar{\gamma}(\bar{\Phi})}\bar{g}_{\mu\nu} \,,	 \\
\label{field.redefinition}
\Phi &= \bar{f}(\bar{\Phi}) \,.	
\end{align}
The first of them is known as the Weyl rescaling, a distinct case of the conformal transformation of the metric tensor $g_{\mu\nu}$ and occasionally we shall refer to it as the change of the `frame'. The second one is the redefinition of the scalar field $\Phi$, also known as `reparametrization'. The transformed action functional reads
\begin{align}
\nonumber
{\bar S} =& \frac{1}{2\kappa^2}\int_{V_4}\mathrm{d}^4x\sqrt{-{\bar g}}
\{ {\bar {\mathcal A}}({\bar \Phi}){\bar R}-{\bar {\mathcal B}}({\bar \Phi})
{\bar g}^{\mu\nu}{\bar \nabla}_\mu{\bar \Phi}{\bar \nabla}_\nu{\bar \Phi} - 2\ell^{-2}{\bar {\mathcal V}}
({\bar \Phi}) \} + \bar{S}_\mathrm{m}\left[\mathrm{e}^{2{\bar \alpha}({\bar \Phi})}{\bar g}_{\mu\nu},\chi^A \right] \\
\label{fl.teisendatud.moju}
& + \frac{1}{2\kappa^2}\int_{V_4}\mathrm{d}^4x\partial_\sigma \left( \sqrt{-\bar{g}} \bar{\mathscr{B}}_{ ( \bar{S} ) }^{\,\sigma} \right)\,
\end{align}
where
\begin{equation}
\label{boundary.term.action}
\bar{\mathscr{B}}_{ ( \bar{S} ) }^{\,\sigma} = - 6\bar{\gamma}^\prime \bar{\mathcal{A}}\bar{g}^{ \sigma\mu}\bar{\nabla}_\mu \bar{\Phi} \,
\end{equation}
is a negligible boundary term. Here we have made use of the following notation \cite{Flanagan}
\begin{subequations}
\label{fl.fnide.teisendused}
\begin{align}
\label{fl.fnide.teisendused:A}
	\bar{\mathcal{A}}(\bar{\Phi}) &= \mathrm{e}^{2\bar{\gamma}(\bar{\Phi})}
	{\mathcal A} \left( {\bar f}( {\bar \Phi})\right) \,,\\
\label{fl.fnide.teisendused:B}
	{\bar {\mathcal B}}({\bar \Phi}) &= \mathrm{e}^{2{\bar \gamma}({\bar \Phi})}\left( 
	\left(\bar{f}^\prime\right)^2{\mathcal B}\left(\bar{f}(\bar{\Phi})\right) -
	6\left(\bar{\gamma}^\prime\right)^2{\mathcal A}\left(\bar{f}(\bar{\Phi})\right) -
	6\bar{\gamma}^\prime\bar{f}^\prime \mathcal{A}^\prime \right) \,, \\
\label{fl.fnide.teisendused:V}
	\bar{{\mathcal V}}(\bar{\Phi}) &= \mathrm{e}^{4\bar{\gamma}(\bar{\Phi})} \, {\mathcal V}\left(\bar{f}(\bar{\Phi})\right) \,, \\
\label{fl.fnide.teisendused:alpha}
	\bar{\alpha}(\bar{\Phi}) &= \alpha\left(\bar{f}(\bar{\Phi})\right) + \bar{\gamma}(\bar{\Phi})\,,
\end{align}
\end{subequations}
and refined the convention \eqref{convention} in order to distinguish between derivatives w.r.t.~the ``barred'' scalar field $\bar{\Phi}$ and the ``unbarred'' scalar field $\Phi$ in the following manner:
\begin{equation}
\bar{f}^\prime  \equiv \frac{\mathrm{d}\bar{f}(\bar{\Phi})}{\mathrm{d}\bar{\Phi}} \,,\qquad \bar{\gamma}^{\,\prime} \equiv \frac{\mathrm{d}\bar{\gamma}(\bar{\Phi})}{\mathrm{d}\bar{\Phi}} \,,\qquad \bar{\mathcal{A}}^\prime \equiv \frac{\mathrm{d}\bar{\mathcal{A}}(\bar{\Phi})}{\mathrm{d}\bar{\Phi}} \,,\qquad \mathcal{A}^\prime \equiv \frac{\mathrm{d} \mathcal{A}(\Phi)}{\mathrm{d} \Phi} \,,\quad \text{etc.}
\end{equation}
If we impose a condition that the action functional \eqref{fl.moju} is invariant under the local Weyl rescaling of the metric tensor \eqref{conformal.transformation} and under the scalar field redefinition \eqref{field.redefinition} then equations \eqref{fl.fnide.teisendused} are the transformation properties of the four arbitrary functions $\left\lbrace \mathcal{A},\,\mathcal{B},\,\mathcal{V},\,\alpha \right\rbrace$. In the current paper we will adopt the aforementioned assumption and whenever the transformations \eqref{conformal.transformation}-\eqref{field.redefinition} are recalled also equations \eqref{fl.fnide.teisendused} are taken into account.

Sometimes it might be more clear to look the transformations also backwards. In order to keep the notation under better control we also introduce
\begin{align}
\label{conformal.transformation.backwards}
\bar{g}_{\mu\nu} &= \mathrm{e}^{2\gamma(\Phi)}g_{\mu\nu}  \,,\\
\label{field.redefinition.backwards}
\bar{\Phi} &= f(\Phi) \,,
\end{align}
such that $\gamma\left( \bar{f}\left(\bar{\Phi}\right) \right) = -\bar{\gamma}\left( \bar{\Phi} \right)$. If $\bar{f}$ is a bijection then the composition $\bar{f}\circ f$ is equal to the identity transformation but we also want to include the possibility that either $\bar{f}$ or $f$ or both are multivalued. When using the transformations \eqref{conformal.transformation.backwards}-\eqref{field.redefinition.backwards} instead of \eqref{conformal.transformation}-\eqref{field.redefinition} then for the transformation rules \eqref{fl.fnide.teisendused} of the four arbitrary functions the property of being ``barred'' or not is interchanged. For an example compare \eqref{A.prime} with \eqref{A.prime.backwards}. 

In the literature most of the calculations have been carried out in a specific parametrization, e.g.~in JF BDBW parametrization \eqref{Jordan.frame} or in EF canonical parametrization \eqref{Einstein.frame}. A specific parametrization is in principle equivalent to the general one \cite{Faraoni:no:progress,JKSV:2} but it turns out that for specific parametrizations the transformation from one to another may not be so unique at all since there are quantities that in these parametrizations remain unseen but nevertheless have complicated transformation rules \cite{JKSV:2,JKSV:1}. As an example let us consider $\mathcal{A}^\prime$ in JF BDBW parametrization. We obtain $\left. \mathcal{A}^\prime \right|_\mathrm{J} = 1$. Hence an arbitrary power of the latter is also equal to one and in that specific parametrization we cannot distinguish between $\left. \mathcal{A}^\prime \right|_\mathrm{J}$, $\left( \left. \mathcal{A}^\prime \right|_\mathrm{J} \right)^2$, etc. However all these have different transformation properties. In the current paper, in order to overcome that shortcoming, we have adopted the notation by Flanagan \cite{Flanagan} which has the following advantages: i) all four possible couplings (curvature $(\mathcal{A})$, kinetic $(\mathcal{B})$, self interaction $(\mathcal{V})$ and matter $(\alpha)$) of the scalar field are explicitly written out, ii) two transformation functions $\bar{\gamma}$ and $\bar{f}$ are kept separate from the coupling functions. 

%%%%%%%%%%%%%%%%%%%%%%%%%%%%%%%%%%%%%%%%%%%%%%%%%%%%%%%%%%%%%%%%%%%%%%%%%%%%%%%%%%%%%%%%%%%%%%%%%%%%%%%%%%%%%%%%%%

\subsubsection{Transformation of the equations of motion}\label{TPOEOM}

A straightforward calculation shows that under the local rescaling of the metric tensor \eqref{conformal.transformation} and the scalar field redefinition \eqref{field.redefinition} the equation of motion \eqref{tensor.equation} for the metric tensor $g_{\mu\nu}$, shortly denoted as $E^{(g)}_{\mu\nu} = 0$, transforms as follows
\begin{equation}
\label{transformation.of.tensor.equation}
E^{(g)}_{\mu\nu} = \mathrm{e}^{-2\bar{\gamma}}\bar{E}^{(\bar{g})}_{\mu\nu} \,.
\end{equation}
Here we have made use of the fact that under the conformal transformation \eqref{conformal.transformation} the energy-momentum tensor $T_{\mu \nu}$ transforms as $T_{\mu \nu} = \mathrm{e}^{ -2 \bar{\gamma}} \bar{T}_{\mu \nu}$ and its contraction as $T = \mathrm{e}^{-4 \bar{\gamma}} \bar{T}$.

Checking the transformation properties of the scalar field equation \eqref{scalar.field.equation.containing.R} that explicitly contains $R$ gives
\begin{equation}
\label{transformation.of.scalar.equation.with.R}
E^{(\Phi)} = \left( \bar{f}^\prime \right)^{-1} \mathrm{e}^{-4\bar{\gamma}}\left\lbrace \bar{E}^{(\bar{\Phi})} + 2\bar{\gamma}^\prime \bar{g}^{ \mu\nu } \bar{E}^{ (\bar{g}) }_{\mu\nu} \right\rbrace \,.
\end{equation}
Therefore these transformations mix the scalar field equation \eqref{scalar.field.equation.containing.R} with the metric equation \eqref{tensor.equation}. The reason for that lies in the transformation properties of the variational derivatives
\begin{equation}
\label{transformation.of.variational.derivatives}
\renewcommand*{\arraystretch}{2}
\begin{pmatrix}
\displaystyle \frac{\delta \phantom{\Phi}}{\delta \Phi} \\
\displaystyle \frac{ \delta \phantom{g^{\sigma\rho}} }{ \delta g^{\sigma\rho} }
\end{pmatrix} =
\begin{pmatrix}
\displaystyle \frac{\delta \bar{\Phi}}{\delta \Phi} &  \displaystyle \frac{\delta \bar{g}^{ \mu\nu } }{\delta \Phi } \\
\displaystyle \frac{\delta \bar{\Phi} }{ \delta g^{\sigma\rho} } & \displaystyle \frac{\delta \bar{g}^{\mu\nu} }{\delta g^{\sigma\rho} }
\end{pmatrix}
\begin{pmatrix}
\displaystyle \frac{\delta \phantom{\Phi}}{\delta \bar{\Phi} } \\
\displaystyle \frac{\delta \phantom{g^{\mu\nu}}}{\delta \bar{g}^{\mu\nu} }
	\end{pmatrix} =
\begin{pmatrix}
\displaystyle \left( \bar{f}^{\prime} \right)^{-1} & \displaystyle 2\bar{\gamma}^{\,\prime} \left( \bar{f}^{\prime} \right)^{-1} \bar{g}^{\,\mu\nu} \\
\displaystyle 0 & \displaystyle \mathrm{e}^{2\bar{\gamma}} \delta^\mu_\sigma \delta^\nu_\rho
\end{pmatrix}
\begin{pmatrix}
\displaystyle \frac{\delta \phantom{\Phi}}{\delta \bar{\Phi} } \\
\displaystyle \frac{\delta \phantom{g^{\mu\nu}}}{\delta \bar{g}^{\mu\nu} }
\end{pmatrix} \,.
\end{equation}
\renewcommand*{\arraystretch}{1}
In the context of the transformations \eqref{conformal.transformation}-\eqref{field.redefinition} the prescription for using the contraction $g^{\mu\nu} E^{(g)}_{\mu\nu}$ to eliminate $R$ from the scalar field equation of motion \eqref{scalar.field.equation.containing.R} can be seen as giving an unconfounded equation under the transformation. Namely
\begin{equation}
\label{transformation.of.scalar.field.equation.without.R}
E^{(\Phi)} + \frac{\mathcal{A}^\prime}{\mathcal{A}}g^{\mu\nu}E^{(g)}_{\mu\nu}= \mathrm{e}^{-4\bar{\gamma}} \left(\bar{f}^\prime\right)^{-1} \left\lbrace \bar{E}^{(\bar{\Phi})} + \frac{\bar{\mathcal{A}}^\prime}{\bar{\mathcal{A}}}\bar{g}^{\mu\nu}\bar{E}^{(\bar{g})}_{\mu\nu} \right\rbrace   \,.
\end{equation}
Note that as under transformations $\displaystyle \frac{\delta \phantom{\Phi}}{ \delta \Phi }$ gains an additive term also $\displaystyle \delta g^{\mu\nu}$ gains one which of course follows from \eqref{conformal.transformation}. Since the action functional $S_\mathrm{m}$ for the matter fields $\chi^A$ functionally depends on $\mathrm{e}^{2\alpha}g_{\mu\nu}$ which is invariant under the transformations \eqref{conformal.transformation}-\eqref{field.redefinition} in the sense of Subsec.~\ref{Invariants} \cite{Flanagan,JKSV:2} it follows that also the equations of motion \eqref{matter.eom} for the matter fields are invariant under these transformations. In order to sum up let us take a look at the transformation of the varied action \eqref{varied.action}. A straightforward calculation reveals 
\begin{align}
\nonumber
\delta S =& \,\frac{1}{2\kappa^2}\int_{V_4} \mathrm{d}^4x  \sqrt{-g} \left\lbrace E^{(g)}_{\mu\nu} \delta g^{\mu\nu} + E^{(\Phi)} \delta\Phi + 2\kappa^2 \mathrm{e}^{4\alpha} E^{(\chi)}_{A} \delta\chi^{A} \right\rbrace \\
\nonumber
&+ \frac{1}{2\kappa^2}\int_{V_4} \mathrm{d}^4x\partial_\sigma \left( \sqrt{-g}\left[ \mathscr{B}_{(g)}^{\,\sigma} + \mathscr{B}_{(\Phi)}^{\,\sigma} + 2 \kappa^2 \mathrm{e}^{4\alpha} \mathscr{B}_{(\chi)}^{\,\sigma} \right]\right)  \\
\nonumber
=& \,\frac{1}{2\kappa^2}\int_{V_4} \mathrm{d}^4x  \sqrt{-\bar{g}} \left\lbrace \bar{E}^{(\bar{g})}_{\mu\nu} \left(  \delta \bar{g}^{\mu\nu} - 2 \bar{\gamma}^{\,\prime} \bar{g}^{\mu\nu} \delta \bar{\Phi} \right) + \left(  \bar{E}^{(\bar{\Phi})} + 2\bar{\gamma}^\prime \bar{g}^{ \mu\nu } \bar{E}^{ (\bar{g}) }_{\mu\nu} \right) \delta \bar{\Phi} + 2\kappa^2 \mathrm{e}^{4\bar{\alpha}} \bar{E}^{(\chi)}_{A} \delta\chi^{A} \right\rbrace \\ 
\nonumber
& + \frac{1}{2\kappa^2}\int_{V_4} \mathrm{d}^4x \partial_\sigma \left( \sqrt{-\bar{g}}\left[ \bar{\mathscr{B}}_{(\bar{g})}^{\,\sigma} + \bar{\mathscr{B}}_{(\bar{\Phi})}^{\,\sigma} \right] + \delta \left( \sqrt{ -\bar{g} }\, \bar{\mathscr{B}}^{\,\sigma}_{ ( \bar{S} ) } \right) +  \sqrt{-\bar{g}} \, 2 \kappa^2 \mathrm{e}^{4\bar{\alpha}} \bar{\mathscr{B}}_{(\chi)}^{\,\sigma} \right)\, \\
=&\, \frac{1}{2\kappa^2}\int_{V_4} \mathrm{d}^4x  \sqrt{-\bar{g}} \left\lbrace \bar{E}^{(\bar{g})}_{\mu\nu} \delta \bar{g}^{\mu\nu}  +  \bar{E}^{(\bar{\Phi})} \delta \bar{\Phi} + 2\kappa^2 \mathrm{e}^{4\bar{\alpha}} \bar{E}^{(\chi)}_{A} \delta\chi^{A} \right\rbrace + \overline{\mathscr{B}\textit{oundary terms} } \,.
\end{align}
The fourth line forms as follows: under the transformations \eqref{conformal.transformation}-\eqref{field.redefinition} the boundary terms \eqref{boundary.term.metric} and \eqref{boundary.term.scalar.field} mix with each other and some extra terms arise. The latter are exactly the ones obtained by varying the boundary term \eqref{boundary.term.action} which arose due to rewriting the action functional \eqref{fl.moju} in terms of $\bar{g}_{\mu\nu}$ and $\bar{\Phi}$. The boundary terms $\bar{\mathscr{B}}^{\,\sigma}_{(\chi)}$ that appear when the action functional \eqref{fl.moju} is varied w.r.t.~the matter fields $\chi^A$ are invariant. As before, we have included the boundary terms for the sake of completeness although they do not contribute to the equations of motion. Indeed, from the viewpoint of the transformation properties also they must behave well.

One can think about the continuity equation \eqref{EI.jaavus} in the same spirit. Let us consider a symmetric second order tensor $E_{(\mu\nu)}$ having the following transformation properties: $E_{(\mu\nu)} = \mathrm{e}^{-2\bar{\gamma}} \bar{E}_{(\mu\nu)}$. For such tensor
\begin{equation}
\label{transformation.of.divergence}
\nabla^\mu E_{\mu\nu} = \mathrm{e}^{ -4\bar{\gamma}}  \bar{ \nabla }^{\mu} \bar{E}_{\mu\nu} - \bar{\gamma}^{\,\prime} \mathrm{e}^{-4\bar{\gamma}}  \bar{g}^{\mu\lambda} \bar{E}_{\mu\lambda} \bar{\nabla}_\nu \bar{\Phi}
\end{equation}
holds. A straightforward calculation shows that the previous knowledge is at least implicitly taken into account when the continuity equation is constructed. Indeed, by making use of \eqref{transformation.of.divergence}, the transformation properties \eqref{transformation.of.tensor.equation} of the tensor equation \eqref{tensor.equation} and \eqref{transformation.of.scalar.equation.with.R} covering the transformation properties of the scalar field equation \eqref{scalar.field.equation.containing.R}, we obtain
\begin{align}
\nonumber
E^{(c)}_{\nu} &\equiv \nabla^\mu E^{(g)}_{\mu\nu} + \frac{1}{2} E^{(\Phi)} \nabla_\nu\Phi = \mathrm{e}^{ -4\bar{\gamma}} \left\lbrace  \bar{ \nabla }^{\mu} \bar{E}^{(\bar{g})}_{\mu\nu} \negmedspace - \negmedspace \bar{\gamma}^{\,\prime} \bar{g}^{\mu\lambda} \bar{E}^{ (\bar{g}) }_{\mu\lambda} \bar{\nabla}_\nu \bar{\Phi} + \frac{1}{2} \left( \bar{E}^{(\bar{\Phi})} \negmedspace + \negmedspace 2\bar{\gamma}^\prime \bar{g}^{ \mu\lambda } \bar{E}^{ (\bar{g}) }_{\mu\lambda} \right) \bar{\nabla}_\nu \bar{\Phi} \right\rbrace \\
\label{transformation.of.continuity.equation}
&= \mathrm{e}^{ -4\bar{\gamma}}\left\lbrace  \bar{ \nabla }^{\mu} \bar{E}^{(\bar{g})}_{\mu\nu} + \frac{1}{2} \bar{E}^{(\bar{\Phi})} \bar{\nabla}_\nu \bar{\Phi} \right\rbrace = \mathrm{e}^{ -4\bar{\gamma}} \bar{\rule[1.7ex]{-0.4ex}{0ex}E}^{(c)}_{\nu} \,.
\end{align}

Hence we have equations of motion given by \eqref{tensor.equation}, \eqref{matter.eom}, \eqref{scalar.field.equation.without.R} and \eqref{EI.jaavus} which only gain a common multiplier under the local Weyl rescaling of the metric tensor \eqref{conformal.transformation} and under the scalar field redefinition \eqref{field.redefinition} but otherwise preserve their structure. An analogous conclusion was drawn in \cite{Morris}. We deem that as these are general equations no problems arise when either the transformation \eqref{conformal.transformation} or \eqref{field.redefinition} become singular at some isolated scalar field value.

%%%%%%%%%%%%%%%%%%%%%%%%%%%%%%%%%%%%%%%%%%%%%%%%%%%%%%%%%%%%%%%%%%%%%%%%%%%%%%%%%%%%%%%%%%%%%%%%%%%%%%%%

\subsubsection{Transformations in the Friedmann cosmology}

Previously the transformation properties of the field equations were discussed. The Friedmann cosmology is a particular case and the corresponding equations of motion \eqref{First.Friedmann.general.equation}-\eqref{cosmology.continuity.equation} transform according to the rules \eqref{transformation.of.tensor.equation}, \eqref{transformation.of.scalar.field.equation.without.R} and \eqref{transformation.of.continuity.equation}, of course. Nevertheless there are some details that need to be mentioned. The line element in Friedmann cosmology has the form \eqref{FLRW}. In order to keep that form of the metric each conformal transformation $g_{\mu\nu}=\mathrm{e}^{2\bar{\gamma}}\bar{g}_{\mu\nu}$ is followed by a time coordinate transformation and a redefinition of the scale factor
\begin{equation}
\label{time.coordinate.transformation}
\mathrm{d}t \mapsto \mathrm{d}\bar{t}:\,\sqrt{\mathrm{e}^{2\bar{\gamma}}}\mathrm{d}\bar{t} = \mathrm{d}t \,; \qquad a(t) \mapsto \bar{a}(\bar{t}):\,\sqrt{\mathrm{e}^{2\bar{\gamma}}}\bar{a}(\bar{t}) = a(t)\,.
\end{equation}
Therefore as the cosmological time depends on the chosen parametrization we adopt the following notation
\begin{equation}
\dot{\left(\right)} \equiv \frac{\mathrm{d}\phantom{t}}{\mathrm{d}t}\left(\right) \text{ and } \dot{\bar{()}} \equiv \frac{\mathrm{d}\phantom{t}}{\mathrm{d}\bar{t}}\bar{()} \,.
\end{equation} 

Due to \eqref{time.coordinate.transformation} the transformation of the Hubble parameter reads
\begin{equation}
\label{transformation.of.H}
H \equiv \frac{\dot{a}}{a} = \mathrm{e}^{-\bar{\gamma}} \left( \bar{H} + \bar{\gamma}^{\prime} \dot{\bar{\Phi}} \right) \,.
\end{equation}
One can counter the additive term arising in \eqref{transformation.of.H} for example by considering the quantity
\begin{equation}
\left(H + \frac{1}{2} \frac{\mathcal{A}^\prime}{\mathcal{A}} \dot{\Phi} \right) = \mathrm{e}^{-\bar{\gamma}}\left(\bar{H} + \frac{1}{2}\frac{\bar{\mathcal{A}}^\prime}{\bar{\mathcal{A}}}\dot{\bar{\Phi}}\right) \,.
\end{equation}
Note also that
\begin{equation}
\label{transformation.of.Phi.dot}
\dot{\Phi} = \mathrm{e}^{-\bar{\gamma}}\bar{f}^\prime\dot{\bar{\Phi}} \,,\qquad \ddot{\Phi} = \mathrm{e}^{-2\bar{\gamma}}\left( \bar{f}^\prime \ddot{\bar{\Phi}} + \bar{f}^{\prime\prime}\dot{\bar{\Phi}}^2 - \bar{\gamma}^{\,\prime}\bar{f}^\prime \dot{\bar{\Phi}}^2 \right)
\end{equation}
are respectively the transformations of the first and the second derivative of the scalar field w.r.t.~the cosmological time.

The transformation of $\rho$ and $p$ is determined by the transformation of the contraction $T$ of the energy-momentum tensor $T_{\mu\nu}$. Using the transformation rule \eqref{transformation.of.variational.derivatives} of the variational derivative $\frac{\delta \phantom{g^{\mu\nu}} }{\delta g^{\mu\nu}}$ on the definition \eqref{energy.momentum.tensor} of the matter energy-momentum tensor reveals that $T = \mathrm{e}^{-4\bar{\gamma}}\bar{T}$ as also mentioned after \eqref{transformation.of.tensor.equation}.

%%%%%%%%%%%%%%%%%%%%%%%%%%%%%%%%%%%%%%%%%%%%%%%%%%%%%%%%%%%%%%%%%%%%%%%%%%%%%%%%%%%%%%%%%%%%%%%%%%%%%%%%%%%%%%%%%%%%

\subsection{Invariants}\label{Invariants}

A more closer look to the transformation rules \eqref{fl.fnide.teisendused} of the four arbitrary functions $\left\lbrace \mathcal{A},\, \mathcal{B},\, \mathcal{V},\, \alpha \right\rbrace$ allows us to write out objects that do not gain any additive or multiplicative terms under the local Weyl rescaling \eqref{conformal.transformation} and under the scalar field redefinition \eqref{field.redefinition}. Let us recall the three basic ones introduced in our recent paper \cite{JKSV:2}
\begin{equation}
\label{invariants}
\mathcal{I}_1(\Phi) \equiv \frac{\mathrm{e}^{2\alpha(\Phi)}}{\mathcal{A}(\Phi)} \,,\quad
\mathcal{I}_2(\Phi) \equiv \frac{\mathcal{V}(\Phi)}{\left(\mathcal{A}(\Phi)\right)^2} \,,\quad
\mathcal{I}_3(\Phi) \equiv \pm\int\sqrt{\mathcal{F}(\Phi)}\mathrm{d}\Phi \,
\end{equation}
where \cite{Flanagan}
\begin{equation}
\label{definition.of.F}
\mathcal{F} \equiv \frac{2 \mathcal{A}\mathcal{B} + 3\left( \mathcal{A}^\prime \right)^2}{4\mathcal{A}^2} \,,\quad \bar{\mathcal{F}} = \left( \bar{f}^\prime \right)^2 \mathcal{F} \,.
\end{equation}
Under the scalar field redefinition \eqref{field.redefinition} these quantities transform as scalar functions but their numerical value at some spacetime point $x^{\mu} \in V_4$ is nevertheless invariant. One can introduce further objects having the same transformation properties by making use of three operations: i) forming an arbitrary function of the invariants \eqref{invariants} etc., ii) introducing a quotient of derivatives $ \mathcal{I}_i \equiv \mathcal{I}_j^\prime/\mathcal{I}_k^\prime$ or iii) integrating the previous result $\mathcal{I}_j \equiv \int \mathcal{I}_i(\Phi) \mathcal{I}_k^\prime(\Phi) \mathrm{d}\Phi$ in the sense of an indefinite integral \cite{JKSV:2}.

The basic quantities \eqref{invariants} were chosen since they are well known and used in the literature. For instance in JF BDBW parametrization $\mathcal{I}_2 = \mathcal{V}_\mathrm{J}/\Psi^2$ and $\mathcal{I}_2^\prime = \left( \Psi \mathcal{V}_\mathrm{J}^\prime - 2 \mathcal{V}_\mathrm{J} \right)/\Psi^3$ determines the fixed points in \cite{Faraoni:Jensen:Theuerkauf}, while in \cite{Toporenski} the term `effective potential' refers to $\mathcal{I}_2$. The invariant $\mathcal{I}_3$ is essential in the Barrow and Parsons solution generating prescription \cite{Barrow:Parsons}. Last but not least in JF BDBW parametrization $1/\mathcal{I}_1=\Psi$ and in EF canonical parametrization $\pm\mathcal{I}_3 = \varphi + const$. Therefore 
\begin{equation}
	\frac{\mathrm{d}\frac{1}{\mathcal{I}_1}}{\mathrm{d} \mathcal{I}_3} = -\frac{1}{\mathcal{I}_1^2} \frac{\mathcal{I}_1^\prime}{\mathcal{I}_3^\prime} = \pm \frac{\mathrm{d}\Psi}{\mathrm{d}\varphi}
\end{equation}
can also considered to be an invariant.

%%%%%%%%%%%%%%%%%%%%%%%%%%%%%%%%%%%%%%%%%%%%%%%%%%%%%%%%%%%%%%%%%%%%%%%%%%%%%%%%%%%%%%%%%%%%%%%%%%%%%%%%%%%%%%%%%%%%%%%%%%%%%%%%%%%%%%%%%%%%%%%%%%%%%%%%%%%%

\section{General relativity regime}\label{general.relativity.regime}

In this section we first write down the conditions under which STG coincides with GR, i.e.~we introduce the notion of the `GR regime'. Second we consider the GR limit, i.e.~a dynamical approach to the GR regime.

%%%%%%%%%%%%%%%%%%%%%%%%%%%%%%%%%%%%%%%%%%%%%%%%%%%%%%%%%%%%%%%%%%%%%%%%%%%%%%%%%%%%%%%%%%%%

\subsection{Theory: part II}\label{subsec:general.relativity.regime}

%%%%%%%%%%%%%%%%%%%%%%%%%%%%%%%%%%%%%%%%%%%%%%%%%%%%%%%%%%%%%%%%%%%%%%%%%%%%%%%%%%%%%%%%%%%%

\subsubsection{General relativity regime}

GR is in a rather good agreement with the experiments carried out in the solar system. Therefore whatever theory of gravitation we consider its predictions, in order to be viable, must be close to those of GR at least in the sufficient neighbourhood of the sun. In the current paper we will bestow consideration upon STG in which the predictions are close to the ones obtained from GR because the field equations themselves are the same at least in some regime. We shall use `GR regime' to refer to such a situation.

In the Einstein's GR the tensor equation, a specific case of \eqref{tensor.equation}, does not contain the terms $\mathcal{B}\nabla_\mu \Phi \nabla_\nu \Phi$, $\mathcal{A}^{\prime\prime}\nabla_\mu \Phi \nabla_\nu \Phi$, $\mathcal{A}^\prime \nabla_\mu \nabla_\nu \Phi$ or the contractions of these. Requiring that $\mathcal{B}$ and the derivatives of $\mathcal{A}$ are zero at the same value of the scalar field $\Phi$ in a generic theory needs finetuning and therefore we instead impose that in the GR regime the scalar field is constant $\Phi = \Phi_0$, i.e.
\begin{equation}
\label{vanishing.nabla.Phi}
\left.\nabla_\mu \Phi \right|_{\Phi_0} = 0
\end{equation}
and $\nabla_\mu\nabla_\nu \Phi = 0$. In this case \eqref{tensor.equation} reduces to the Einstein equation in GR, with $\kappa^2/\mathcal{A}(\Phi_0)$ playing the role of the gravitational constant and $\ell^{-2}\mathcal{V}(\Phi_0)$ as the cosmological constant, both positive. Also the continuity equation \eqref{EI.jaavus} reduces to $\nabla^\mu T_{\mu\nu} = 0$. 

In order to maintain the constancy of the scalar field $\Phi$ the equation of motion \eqref{scalar.field.equation.without.R} must become an identity $0 = 0$ at the scalar field value $\Phi_0$. Let us divide \eqref{scalar.field.equation.without.R} by $4\mathcal{A}\mathcal{F}$ and make use of the invariant objects \eqref{invariants} and $\mathcal{F}$, given by \eqref{definition.of.F}, in order to rewrite \eqref{scalar.field.equation.without.R} in a more compact manner as follows
\begin{equation}
\label{scalar.field.equation.without.R.divided.by.F}
\Box \Phi + \left( \frac{\mathcal{A}^\prime}{\mathcal{A}} - \frac{1}{2}\left( \frac{1}{\mathcal{F}} \right)^\prime \mathcal{F} \right) g^{\mu\nu} \nabla_\mu\Phi \nabla_\nu\Phi = \frac{\mathcal{A}}{2\ell^2}\frac{ \mathcal{I}_2^\prime }{ \mathcal{F} } - \frac{\kappa^2}{ 4\mathcal{A} } \frac{\left( \ln \mathcal{I}_1 \right)^\prime }{ \mathcal{F} } T \,.
\end{equation}
The l.h.s.~of \eqref{scalar.field.equation.without.R.divided.by.F} contains derivatives and therefore for a constant scalar field value it vanishes. Hence in order to avoid finetuning we impose that also the source terms
\begin{align}
\label{potential.condition}
\left.\frac{\mathcal{A}}{2\ell^2}\frac{ \mathcal{I}_2^\prime }{ \mathcal{F} }\right|_{\Phi_0} &= \left.2\ell^{-2} \frac{\left( \mathcal{A}\mathcal{V}^\prime \negmedspace - \negmedspace 2\mathcal{V}\mathcal{A}^\prime  \right)}{2\mathcal{A}\mathcal{B} \negmedspace + \negmedspace 3\left( \mathcal{A}^\prime\right)^2} \right|_{\Phi_0} = 0 \,, \\
\label{matter.condition}
\left.\frac{\kappa^2}{ 4\mathcal{A} } \frac{\left( \ln \mathcal{I}_1 \right)^\prime }{ \mathcal{F} } \right|_{\Phi_0} &= \left. \kappa^2 \frac{\left( 2\mathcal{A}\alpha^\prime \negmedspace - \negmedspace \mathcal{A}^\prime \right)}{2\mathcal{A}\mathcal{B} \negmedspace + \negmedspace 3\left( \mathcal{A}^\prime\right)^2} \right|_{\Phi_0} = 0\,
\end{align}
in the regime where the predictions of the theory described by the action functional \eqref{fl.moju} are close to those of GR. In the following we shall use `vanishing source conditions' for referring to \eqref{potential.condition}-\eqref{matter.condition}. In the JF BDBW parametrization \eqref{Jordan.frame} the second condition \eqref{matter.condition} can only be satisfied by letting $\omega(\Psi) \to \infty$ \cite{JKSV:1} and in that case also the first condition is satisfied.

Let us point out that one may also consider a situation where on the r.h.s.~of \eqref{scalar.field.equation.without.R.divided.by.F} the sum vanishes but both additive terms separately are nonvanishing. In this case a so called screening mechanism is operating, e.g.~the chameleon effect \cite{Khoury:Weltman} or the symmetron screening mechanism \cite{Hinterbichler:Khoury}. However in these cases vanishing of the r.h.s.~of \eqref{scalar.field.equation.without.R.divided.by.F} depends on the matter contribution. If the latter changes, e.g.~the energy density $\rho$ in the Friedmann cosmology decreases as the universe expands, then also the scalar field must evolve further. In the current paper we are interested in basic cosmological scenarios where the scalar field dynamics ends once and for all and therefore we do not focus upon the screening mechanisms. 

For a specific matter content with $T \equiv 0$, e.g.~radiation, the condition \eqref{matter.condition} is not needed \cite{JKSV:1}. If in addition to the latter also $\mathcal{V} \equiv 0$ is considered then the r.h.s.~of \eqref{scalar.field.equation.without.R.divided.by.F} vanishes automatically, and the GR regime can in principle be realized at any value of $\Phi$. In the context of the original Brans-Dicke theory with a constant parameter $\omega$, i.e.~a particular case of JF BDBW parametrization \eqref{Jordan.frame}, there is a discussion in the literature that in the case of $T \equiv 0$, $\mathcal{V} \equiv 0$ taking parametrically $\omega = \infty$ does not reduce the STG solutions to the ones of GR \cite{Faraoni:1998:etal}. However, if $\nabla_\mu \Psi = 0$ is not imposed, only letting $\omega$ to diverge is not sufficient for obtaining the GR regime indeed.

In addition to the vanishing source conditions \eqref{potential.condition}-\eqref{matter.condition} a little more is needed to achieve GR-like behaviour. Namely, for GR the well known relation $-R + \left( 4\Lambda \right) \propto T$ holds. Let us make use of the latter and obtain some restrictions from the contraction \eqref{contraction.of.tensor.equation} of the tensor equation \eqref{tensor.equation}. First, a short glimpse reveals that vanishing or diverging $\left.\mathcal{A}\right|_{\Phi_0}$ violates the mentioned condition. Second, as $\left.\frac{ \kappa^2 }{ \mathcal{A} }\right|_{\Phi_0}$ is the effective gravitational ``constant'' we impose $\left. \mathcal{A} \right|_{\Phi_0} > 0$ in order to have an attractive gravity. Third, the same equation reveals that the potential $\mathcal{V}$, which at that constant scalar field value $\Phi_0$ mimics the cosmological constant $\Lambda$, must be nondiverging as well. In the current paper we also assume it to be nonnegative. Last but not least we impose that $\alpha$ must be nondiverging because otherwise the coupling of the matter fields to the geometry determined by $g_{\mu\nu}$ is unphysical. These assumptions are below spelled out in \eqref{conditions.on.A.etc}.

Analogously to the previous let us point out that in the context of the GR regime the scalar field equation of motion \eqref{scalar.field.equation.containing.R} containing $R$ might be a constraint equation. We start by assuming $\left| \mathcal{A}^\prime \right|_{\Phi_0} < \infty$ because otherwise the behaviour of the effective gravitational ``constant'' $\frac{\kappa^2}{\mathcal{A}}$ becomes unnatural if the scalar field $\Phi$ deviates from its constant value $\Phi_0$. Under this assumption \eqref{scalar.field.equation.containing.R} reveals that also $\left| \alpha^\prime \right|_{\Phi_0} < \infty$ and $\left| \mathcal{V}^\prime \right|_{\Phi_0} < \infty$ because otherwise the constraint $-R + \left(4 \Lambda \right) \propto T$ is violated. These conditions are captured below as \eqref{conditions.on.A.prime.etc}. For the latter we have implicitly assumed that neither $\mathcal{B}$ nor its derivative diverges. In the current paper we restrict our analysis to the cases where only one out of the four arbitrary functions $\left\lbrace \mathcal{A}\,,\, \mathcal{B}\,,\, \mathcal{V}\,,\, \alpha \right\rbrace$ along with its derivatives might diverge. Hence if diverging $\mathcal{B}$ is under consideration then all other functions are assumed to be regular and therefore \eqref{conditions.on.A.prime.etc} is imposed as a general condition in the GR regime.

Let us analyze the possibilities to satisfy the vanishing source conditions \eqref{potential.condition}-\eqref{matter.condition} in more detail. The first and most obvious one is to demand that both numerators are zero at the same scalar field value $\Phi_0$. The other possibility is to let the denominator diverge at the scalar field value $\Phi_0$. In some sense this is a more natural one because no tuning is needed, i.e.~if one of the conditions is satisfied then the other must be satisfied as well. Since the diverging $\left.\mathcal{A}\right|_{\Phi_0}$ and $\left. \mathcal{A}^\prime \right|_{\Phi_0}$ cases have already been omitted, we are left with possibly diverging $\left. \mathcal{B} \right|_{\Phi_0}$ (i.e.~in essence JF BDBW $\omega(\Psi)$). In what follows we keep the latter in mind but nevertheless make most of the statements about $\mathcal{F}$ where $\mathcal{B}$ resides because the transformation property of $\mathcal{F}$, given by \eqref{definition.of.F}, is remarkably simpler than the rule for $\mathcal{B}$, given by \eqref{fl.fnide.teisendused:B}.

To sum up, we consider two possibilities to fulfil the vanishing source conditions \eqref{potential.condition}-\eqref{matter.condition} \cite{JKSV:2}:
\begin{align}
\label{moon.critical.point}
i)\quad \left\lbrace \Phi_\bullet \right\rbrace &\equiv \left\lbrace 
\Phi \left| \,\, \mathcal{I}_2^\prime = 0 = \mathcal{I}_1^\prime  \text{ and } \frac{1}{\left( \mathcal{I}_3^\prime \right)^2} \equiv \frac{1}{\mathcal{F}} \neq 0 \right. \right\rbrace \,,\\
ii)\quad \label{star.critical.point}
\left\lbrace \Phi_\star \right\rbrace &\equiv \left\lbrace \Phi \left| \,\, \frac{1}{\left( \mathcal{I}_3^\prime \right)^2} \equiv  \frac{1}{\mathcal{F}}  = 0 \right. \right\rbrace \,.
\end{align}
If $\mathcal{F}$ diverges then one must also demand that in the GR regime the last term on the l.h.s.~of \eqref{scalar.field.equation.without.R.divided.by.F} vanishes nevertheless, but this is rather a question about the permitted behaviour of a specific solution, i.e.~the order of magnitude of $\nabla_\mu\Phi \to 0$ w.r.t.~$\Phi - \Phi_0 \to 0$, which we shall not yet discuss.

%%%%%%%%%%%%%%%%%%%%%%%%%%%%%%%%%%%%%%%%%%%%%%%%%%%%%%%%%%%%%%%%%%%%%%

\subsubsection{General relativity limit}\label{subsec:Dynamical.approach.to.the.GR.regime}

Once we have a consistent notion of the GR regime it is of course important to find whether a solution under consideration converges to that regime or repels from it. One useful tool for clarifying the question is provided by the dynamical systems method. In Sec.~\ref{dynamical.system} of the current paper we benefit from that method because the GR regime can be identified with a critical point in the $\left( \Phi\,,\,\nabla_\mu\Phi \right)$ space. More precisely we linearize  \eqref{scalar.field.equation.without.R.divided.by.F}, i.e.~the scalar field equation of motion. According to the Hartman-Grobman theorem, the linearized equation captures the qualitative behaviour of the full dynamics if and only if the critical point is hyperbolic, i.e.~all eigenvalues have nonzero real part. It can be shown that a necessary condition for the critical point to be hyperbolic is given by either of the conditions \cite{JKS2010,JKS2012}
\begin{align}
\label{condition.for.C.2}
0 < \left| C_2 \right| < \infty :\qquad & C_2 \equiv -\left. \frac{\mathrm{d}\phantom{\Phi}}{\mathrm{d}\Phi}\left( \frac{\mathcal{A}}{2\ell^2}\frac{ \mathcal{I}_2^\prime }{ \mathcal{F} } \right)\right|_{\Phi_0} = -\left. \frac{\mathcal{A}}{2\ell^2}\left( \frac{\mathcal{I}_2^{\prime\prime} }{\mathcal{F}} + \left( \frac{1}{\mathcal{F}} \right)^\prime \mathcal{I}_2^{\prime}  \right) \right|_{\Phi_0} \,,\\ 
\label{condition.for.C.3}
0 < \left| C_3 \right| <\infty : \qquad & C_3 \equiv \left. \frac{\mathrm{d}\phantom{\Phi} }{\mathrm{d}\Phi} \left( \frac{\kappa^2}{ 4\mathcal{A} } \frac{\left( \ln \mathcal{I}_1 \right)^\prime }{ \mathcal{F} } \right)\right|_{\Phi_0} = \left. \frac{\kappa^2}{ 4\mathcal{A} } \left( \frac{ \left( \ln \mathcal{I}_1 \right)^{\prime\prime} }{ \mathcal{F}} + \left( \frac{1}{\mathcal{F}} \right)^\prime \left( \ln \mathcal{I}_1 \right)^\prime \right) \right|_{\Phi_0} \,.
\end{align}
In other words we assume that the leading term in the Taylor expansion of the r.h.s.~of \eqref{scalar.field.equation.without.R.divided.by.F} is linear w.r.t.~$\Phi-\Phi_0$. In what follows we shall refer to \eqref{condition.for.C.2}-\eqref{condition.for.C.3} as `first order small source conditions'. In \eqref{condition.for.C.2}-\eqref{condition.for.C.3} we have made use of the vanishing source conditions \eqref{potential.condition}-\eqref{matter.condition} in order to cancel some additive terms. Actually due to the same conditions only one of the additive terms on the r.h.s.-s of \eqref{condition.for.C.2}-\eqref{condition.for.C.3} can be nonvanishing. Perhaps it is instructive to write out these conditions also in EF canonical parametrization \eqref{Einstein.frame} (cf. \cite{Faraoni:Jensen:Theuerkauf,Leon})
\begin{equation}
\label{Einstein.frame.first.ord.small}
\left. C_2 \right|_{\text{EF can.}} = -\left.\frac{1}{2\ell^2}\mathcal{V}_{E}^{\prime\prime}\right|_{ \varphi_0 } \,,\qquad\qquad \left. C_3 \right|_{\text{EF can.}} = \left. \frac{\kappa^2}{2} \alpha_{E}^{\prime\prime} \right|_{\varphi_0} \,
\end{equation}
and in JF BDBW parametrization \eqref{Jordan.frame} (cf. \cite{Faraoni:Jensen:Theuerkauf,JKS2008})
\begin{align}
\nonumber
\left. C_2 \right|_{\text{JF BDBW}} &= -\left. 2\ell^{-2} \left( \frac{\left(\Psi \mathcal{V}^\prime_\mathrm{J} -2\mathcal{V}_\mathrm{J}\right)^{\prime} }{2\omega + 3} + \left( \frac{1}{2\omega + 3}\right)^{\prime} \left( \Psi \mathcal{V}^\prime_\mathrm{J} - 2\mathcal{V}_\mathrm{J} \right) \right) \right|_{\Psi_0} \,,\\
\label{Jordan.frame.first.ord.small}
\left. C_3 \right|_{\text{JF BDBW}} &= -\left. \kappa^2 \left(\frac{1}{2\omega + 3}\right)^{\prime} \right|_{\Psi_0} \,.
\end{align}
Let us make use of the first order small source conditions \eqref{condition.for.C.2}-\eqref{condition.for.C.3} in order to adopt the following three assumptions on $\mathcal{F}$ \cite{JKS2008,JKSV:2}:
\begin{align}
\label{nonvanishing.F}
0 \leq \left. \frac{1}{\mathcal{F}} \right|_{\Phi_0} &< \infty \,,\\
\label{non.diverging.derivatives.of.1.over.F}
-\infty <  \left(\frac{1}{\mathcal{F}}\right)^{\overset{n \text{-times}}{\overbrace{\prime \dots \prime}}} \Bigg|_{\Phi_0} &\equiv \left. \frac{\mathrm{d}^n\phantom{\Phi}}{\mathrm{d}\Phi^n}\left(\frac{1}{\mathcal{F}}\right) \right|_{\Phi_0} < \infty \,,\\
\label{non.vanishing.first.derivative.of.1.over.F}
\text{if } \Phi_0 \equiv \Phi_\star, \text{ see \eqref{star.critical.point}, i.e.} \left.\frac{1}{\mathcal{F}}\right|_{\Phi_\star} &= 0 \,, \text{ then } \left.\left(\frac{1}{\mathcal{F}}\right)^\prime\right|_{\Phi_\star} \neq 0 \,.
\end{align}
The transformation rule for $\mathcal{F}$, given by \eqref{definition.of.F}, reveals that $\mathcal{F}$ preserves its sign under the local Weyl rescaling \eqref{conformal.transformation} and under the scalar field redefinition \eqref{field.redefinition}. Therefore, if we want to stay connected with EF canonical parametrization \eqref{Einstein.frame}, where $\mathcal{F}_\mathrm{E} = 1$, then in any other parametrization $\mathcal{F}$ must be nonnegative. Here we make one step further by imposing \eqref{nonvanishing.F}, i.e.~assuming $\mathcal{F}$ to be strictly positive in order to avoid the possibility that in the vanishing source conditions \eqref{potential.condition}-\eqref{matter.condition} both numerator and denominator vanish. In the following we shall refer to \eqref{nonvanishing.F} as the `positive $\mathcal{F}$ assumption'. Let us point out that in the JF BDBW parametrization \eqref{Jordan.frame} the limit $\mathcal{F}_\mathrm{J}=0$ corresponds to $\omega = -\frac{3}{2}$. 

The `differentiable $\frac{1}{\mathcal{F}}$ assumption' \eqref{non.diverging.derivatives.of.1.over.F} guarantees that we can handle the possible singularity lying in $\mathcal{F}$.  Last but not least the `nonvanishing $\left(\frac{1}{\mathcal{F}}\right)^\prime$ assumption' \eqref{non.vanishing.first.derivative.of.1.over.F} is a necessary condition for the critical point to be hyperbolic. The latter only applies if $\mathcal{F} \to \infty$. Therefore, e.g.~in EF canonical parametrization \eqref{Einstein.frame} the assumption \eqref{non.vanishing.first.derivative.of.1.over.F} is automatically satisfied since $\mathcal{F}_\mathrm{E}=1$ never diverges. It can be shown that if the nonvanishing $\left( \frac{1}{\mathcal{F}} \right)^\prime$ assumption \eqref{non.vanishing.first.derivative.of.1.over.F} is not fulfilled then one cannot express the JF BDBW parametrization scalar field $\Psi$ as a Taylor expansion of the EF canonical parametrization scalar field $\varphi$ \cite{JKSV:2}. Let us point out that if the condition \eqref{non.vanishing.first.derivative.of.1.over.F} holds then the equation of motion \eqref{tensor.equation} for the metric tensor and the equation of motion \eqref{scalar.field.equation.without.R} for the scalar field converge to the GR regime at the same ``rate''. The latter is determined by $\mathcal{F}g^{\mu\nu}\nabla_\mu\Phi \nabla_\nu \Phi \to 0 $ \cite{JKSV:1}.

In order to sum up let us gather the restrictions on the three arbitrary functions $\left\lbrace \mathcal{A},\, \mathcal{V},\, \alpha \right\rbrace$ while $\mathcal{B}$ is covered by assumptions \eqref{nonvanishing.F}-\eqref{non.vanishing.first.derivative.of.1.over.F}:
\begin{align}
\label{conditions.on.A.etc}
0 < \left.\mathcal{A}\right|_{\Phi_0} < \infty \,, \qquad 0 \leq \left.\mathcal{V}\right|_{\Phi_0} &< \infty \,, \qquad 0 < \left. \mathrm{e}^{2\alpha} \right|_{\Phi_0} < \infty \,, \\
\label{conditions.on.A.prime.etc}
\left| \mathcal{A}^\prime \right|_{\Phi_0} < \infty \,, \qquad \left| \mathcal{V}^\prime \right|_{\Phi_0} &< \infty \,, \qquad \left| \alpha^\prime \right|_{\Phi_0} < \infty \,, \\
\label{conditions.on.A.second.etc}
\left| \mathcal{A}^{\prime\prime} \right|_{\Phi_0} < \infty \,, \qquad \left| \mathcal{V}^{\prime\prime} \right|_{\Phi_0} &< \infty \,, \qquad \left| \alpha^{\prime\prime} \right|_{\Phi_0} < \infty \,, \\
\label{nonvanishing.A.prime.etc}
\text{if } \left.\frac{1}{\mathcal{F}}\right|_{\Phi_0}=0\text{ then either  } \left.\mathcal{A}^\prime \right|_{\Phi_0} \neq 0 &\text{  or  }\left.\mathcal{V}^\prime \right|_{\Phi_0} \neq 0 \text{  or  }\left.\alpha^\prime \right|_{\Phi_0} \neq 0 \,, \\
\label{nonvanishing.I.1.second.etc}
\text{if } \left.\frac{1}{\mathcal{F}}\right|_{\Phi_0} \neq  0 \text{ then either  } \left.\mathcal{I}_1^{\prime\prime} \right|_{\Phi_0} \neq 0 &\text{  or  }\left.\mathcal{I}_2^{\prime\prime} \right|_{\Phi_0} \neq 0
\end{align}
where \eqref{conditions.on.A.etc}-\eqref{conditions.on.A.prime.etc} are necessary for a consistent notion of the GR regime and \eqref{conditions.on.A.second.etc}-\eqref{nonvanishing.I.1.second.etc} complement the assumptions \eqref{nonvanishing.F}-\eqref{non.vanishing.first.derivative.of.1.over.F} on $\mathcal{F}$ in order to obtain a hyperbolic critical point when the dynamical systems method is used.

%%%%%%%%%%%%%%%%%%%%%%%%%%%%%%%%%%%%%%%%%%%%%%%%%%%%%%%%%%%%%%%%%%%%%%

\subsubsection{Two remarks}\label{Two.remarks}

Two comments about the assumptions \eqref{nonvanishing.F}-\eqref{non.vanishing.first.derivative.of.1.over.F} on $\mathcal{F}$ are in order. First the nonvanishing $\left( \frac{1}{\mathcal{F}} \right)^\prime$ assumption \eqref{non.vanishing.first.derivative.of.1.over.F} imposes that $\left.\frac{1}{\mathcal{F}}\right|_{\Phi_\star} = 0$ is not an extrema of the same function. Therefore if the scalar field $\Phi$ evolves through the value $\Phi_\star$ then $\frac{1}{\mathcal{F}}$ becomes negative hence violating the positive $\mathcal{F}$ assumption \eqref{nonvanishing.F}. We would expect that a consistent theory is endowed with a mechanism that forbids the violation of the condition \eqref{nonvanishing.F}. In other words if $\left.\left( \frac{1}{\mathcal{F}} \right)^\prime\right|_{\Phi_\star}$ is positive (negative) then the scalar field $\Phi$ values permitted by the field equations should not be allowed to be less (more) than the value $\Phi_{\star}$. Essentially the same was pointed out in the context of the Friedmann cosmology where the argumentation was based on the field space dynamics \cite{JKS2010}.

Second, the differentiable $\frac{1}{\mathcal{F}}$ assumption \eqref{non.diverging.derivatives.of.1.over.F} states that the limiting value
\begin{equation}
\label{diverging.F.equivalent.to.one.over.x}
\lim\limits_{\Phi \to \Phi_\star} \left[ \left( \frac{1}{\mathcal{F}} \right)^\prime \mathcal{F} \cdot \left( \Phi - \Phi_\star \right) \right] = M \,
\end{equation} 
holds. Here if $\mathcal{F}$ diverges, then $M$ is the order of the first nonzero derivative $\left(\frac{1}{\mathcal{F}}\right)^{(M)} \neq 0$, otherwise $M=0$. If also the nonvanishing $\left( \frac{1}{\mathcal{F}} \right)^\prime$ assumption \eqref{non.vanishing.first.derivative.of.1.over.F} is applied then we can always replace $\left( \frac{1}{\mathcal{F}} \right)^\prime \mathcal{F}$ by $\frac{1}{ \Phi - \Phi_\star }$ whenever calculating the limiting values in the process where $\mathcal{F}$ diverges. In the following we shall use the assumptions \eqref{nonvanishing.F}-\eqref{non.vanishing.first.derivative.of.1.over.F} on $\mathcal{F}$ and therefore in the current paper $M=0$ or $M=1$ are the two possibilities.

%%%%%%%%%%%%%%%%%%%%%%%%%%%%%%%%%%%%%%%%%%%%%%%%%%%%%%%%%%%%%%%%%%%%%%%%%%%%%%%%%%%%%%%%%%%%%%%%%%%%%%%%%%%%%%%%%%%%%%%%%%%%%%%%

\subsubsection{Barrow-Parsons classes}

The assumptions \eqref{nonvanishing.F}-\eqref{non.vanishing.first.derivative.of.1.over.F} on $\mathcal{F}$ are restrictive but there are many studies which consider a functional form of $\mathcal{F}$ obeying \eqref{nonvanishing.F}-\eqref{non.vanishing.first.derivative.of.1.over.F}. A rather general classification of the possible functional forms of $\mathcal{F}$ in the JF BDBW parametrization \eqref{Jordan.frame} was given by Barrow and Parsons \cite{Barrow:Parsons}. There they constrained the constant powers $\beta_i$ so that the parametrized post-Newtonian conditions $\omega(\Psi) \to \infty$ and $\frac{\omega^\prime(\Psi)}{\omega(\Psi)^3} \to 0$ are satisfied. Here, analogously to \cite{Minazzoli:Hees}, we write out the further necessary restrictions on the Barrow-Parsons classes so that the assumptions \eqref{nonvanishing.F}-\eqref{non.vanishing.first.derivative.of.1.over.F} are satisfied.

\begin{itemize}
	\item[1) ] $\displaystyle\frac{1}{\mathcal{F}_\mathrm{J}} \equiv \frac{4\Psi^2}{2\omega(\Psi) + 3} \propto \Psi^2\left| 1 - \frac{\Psi}{\Psi_\star} \right|^{\beta_1}$\,, \quad $\beta_1>\frac{1}{2}$ \,.
	\begin{itemize}
		\item[i) ] Assumption \eqref{nonvanishing.F} is fulfilled if $\left| \Psi \right| \not\to \infty$. The latter is assured by assumption \eqref{conditions.on.A.etc}.
		\item[ii) ] Assumption \eqref{non.diverging.derivatives.of.1.over.F} is fulfilled if $\beta_1$ is an arbitrary positive integer power.
		\item[iii) ] Assumption \eqref{non.vanishing.first.derivative.of.1.over.F} is fulfilled if $\beta_1 = 1$. 
	\end{itemize}
	Hence we obtain $\displaystyle\frac{1}{\mathcal{F}_\mathrm{J}} \propto \Psi^2\left| 1 - \frac{\Psi}{\Psi_\star} \right|$ fulfils the assumptions \eqref{nonvanishing.F}-\eqref{non.vanishing.first.derivative.of.1.over.F}. Such functional forms have been considered e.g.~in \cite{Serna:2002fj,SernaAlimi}.
	
	\item[2) ] $\displaystyle \frac{1}{\mathcal{F}_\mathrm{J}} \propto \Psi^2\left| \ln\left( \frac{\Psi}{\Psi_\star} \right) \right|^{\beta_2}$\,, \quad $\beta_2 > \frac{1}{2}$ \,.
	\begin{itemize}
		\item[i) ] Assumption \eqref{nonvanishing.F} is fulfilled if $\left| \Psi \right| \not\to \infty$ and $\Psi \not\to 0$.
		\item[ii) ] Assumption \eqref{non.diverging.derivatives.of.1.over.F} is fulfilled if $\beta_2$ is an arbitrary positive integer power.
		\item[iii) ] Assumption \eqref{non.vanishing.first.derivative.of.1.over.F} is fulfilled if $\beta_2 = 1$. 
	\end{itemize}
	Hence we obtain $\displaystyle\frac{1}{\mathcal{F}_\mathrm{J}} \propto \Psi^2 \left| \ln\left( \frac{\Psi}{\Psi_\star} \right) \right|$ fulfils the assumptions \eqref{nonvanishing.F}-\eqref{non.vanishing.first.derivative.of.1.over.F}. Such functional forms have been considered e.g.~in \cite{DamourNordtvedt:1993a,BarrowMimoso}.
	
	\item[3) ]
	$\displaystyle \frac{1}{\mathcal{F}_\mathrm{J}} \propto \Psi^2 \left| 1 - \left( \frac{\Psi}{\Psi_\star} \right)^{\beta_3} \right|$\,, \quad $\beta_3 > 0$ \,.
	\begin{itemize}
		\item[i) ] Assumption \eqref{nonvanishing.F} is fulfilled if $\left| \Psi \right| \not\to \infty$.
		\item[ii) ] Assumption \eqref{non.diverging.derivatives.of.1.over.F} is fulfilled for arbitrary $\beta_3$.
		\item[iii) ] Assumption \eqref{non.vanishing.first.derivative.of.1.over.F} is fulfilled for arbitrary $\beta_3$.
	\end{itemize}
	Hence we obtain $\displaystyle \frac{1}{\mathcal{F}_\mathrm{J}} \propto \Psi^2 \left| 1 - \left( \frac{\Psi}{\Psi_\star} \right)^{\beta_3} \right|$ fulfils the assumptions \eqref{nonvanishing.F}-\eqref{non.vanishing.first.derivative.of.1.over.F} for arbitrary $\beta_3$. Such theories are studied e.g.~by \cite{Modak:etal,nonminimal,JKS:2007}.
\end{itemize}

%%%%%%%%%%%%%%%%%%%%%%%%%%%%%%%%%%%%%%%%%%%%%%%%%%%%%%%%%%%%%%%%%%%%%%%%%%%%%%%%%%%%%%%%%%%%%%%%%%%%%%%%%%%%%%%%%%%%%%%%%%%%%%%%

\subsection{Transformations: part II}

%%%%%%%%%%%%%%%%%%%%%%%%%%%%%%%%%%%%%%%%%%%%%%%%%%%%%%%%%%%%%%%%%%%%%%%%%%%%%%%%%%%%%%%%%%%%%%%%%%%%%%%%%%%%%%%%%%%%%%%%%%%%%%%%

In the current subsection we analyze the transformation properties in the vicinity of the GR regime. In order to simplify the notation we drop an explicit reference to the point of evaluation $\left(\right)|_0$. Let us start by studying the local Weyl rescaling of the metric tensor \eqref{conformal.transformation} and the scalar field redefinition \eqref{field.redefinition}. The former shall be restricted on mathematical grounds but in order to impose conditions on the latter we make use of the assumptions \eqref{nonvanishing.F}-\eqref{non.vanishing.first.derivative.of.1.over.F} on $\mathcal{F}$. A preluding remark concerning the scalar field redefinition \eqref{field.redefinition} is in order. Let us impose the function $\bar{f}(\bar{\Phi})$ to be at least directionally continuous but retain the possibility that the Jacobian $\bar{f}^\prime \equiv \mathrm{d}\Phi/\mathrm{d}\bar{\Phi}$ of this coordinate transformation in the $1$-dimensional field space may be singular or have zeros at some isolated value of the scalar field $\bar{\Phi}$. The latter is motivated by the observation that in the GR regime $\mathcal{F}$ can be singular in some parametrization.

Whenever the consistency between the constraints imposed on the transformation functions $\bar{\gamma}$ and $\bar{f}$ and on the four arbitrary functions $\left\lbrace \mathcal{A}\,,\, \mathcal{B}\,,\, \mathcal{V}\,,\, \alpha \right\rbrace$ is studied we consider having two parametrizations where the assumptions on the four arbitrary functions hold and then check whether the transformation between these two obey the constraints on the transformation functions.

%%%%%%%%%%%%%%%%%%%%%%%%%%%%%%%%%%%%%%%%%%%%%%%%%%%%%%%%%%%%%%%%%%%%%%%%%%%%%%%%%%%%%%%%%%%%%%%%%%%%%%%%%%%%%%%%%%%%%%%%%%%%%%%%

\subsubsection{Constraints on $\bar{\gamma}$}

Hereby we restrict the local Weyl rescaling of the metric tensor \eqref{conformal.transformation} by making mathematical assumptions and analyze how are the resulting constraints related to the restrictions \eqref{conditions.on.A.etc}-\eqref{conditions.on.A.second.etc} imposed on the three arbitrary functions $\left\lbrace \mathcal{A},\, \mathcal{V},\, \alpha \right\rbrace$.

We start by assuming the local Weyl rescaling of the metric tensor \eqref{conformal.transformation} to be regular, i.e.~the function $\bar{\gamma}(\bar{\Phi})$ and its first and second derivative, $\mathrm{d}\bar{\gamma}/\mathrm{d}\bar{\Phi}$ and $\mathrm{d}^2\bar{\gamma}/\mathrm{d}\bar{\Phi}^2$ respectively do not diverge because otherwise we would introduce geometrical singularities via the local rescaling of the metric. Note that this excludes the interesting possibility of ``conformal continuation'' \cite{Bronnikov:2002kf}. However here we are focussed upon the GR regime which cannot be consistent with the conformal continuation anyway. Let us proceed by pointing out a conclusion that follows from introducing the Weyl rescaling and the scalar field redefinition backwards, i.e.\ \eqref{conformal.transformation.backwards}-\eqref{field.redefinition.backwards},
\begin{equation}
\label{conclusion.for.gamma.prime}
-\bar{\gamma}^{\,\prime} \equiv -\frac{\mathrm{d}\bar{\gamma}(\bar{\Phi})}{\mathrm{d}\bar{\Phi}} \equiv \frac{\mathrm{d}\gamma\left( \bar{f}(\bar{\Phi}) \right)}{\mathrm{d}\bar{\Phi}} = \frac{\mathrm{d} \gamma(\Phi) }{\mathrm{d}\Phi}\frac{\mathrm{d}\Phi}{\mathrm{d}\bar{\Phi}} \equiv \gamma^{\,\prime} \cdot \bar{f}^{\prime}\,.
\end{equation}
From assumption $\left|\bar{\gamma}^{\,\prime}\right| < \infty$ we deduce that if $\left|\bar{f}^\prime\right| \rightarrow \infty$ then in the same process $\gamma^{\,\prime} \rightarrow 0$ because otherwise $\bar{\gamma}^{\,\prime}$ would necessarily diverge. Hence as $\left|\bar{f}^\prime\right| \rightarrow \infty$ implies $f^\prime \rightarrow 0$ we conclude that for any transformation where $f^\prime \rightarrow 0$ also $\gamma^{\,\prime} \rightarrow 0$. The latter is a necessary condition. The actual value of the uncertainty $0 \cdot \infty$ depends on the limiting process which we have assumed to give a nondiverging result.

In order to show that the constraints on the transformation function $\bar{\gamma}$ are in accordance with the assumptions on the three arbitrary functions $\left\lbrace \mathcal{A},\, \mathcal{V},\, \alpha \right\rbrace$, given by \eqref{conditions.on.A.etc}-\eqref{conditions.on.A.second.etc}, let us write out the following:
\begin{align}
\label{A}
\bar{\mathcal{A}} &= \mathrm{e}^{2\bar{\gamma}}\mathcal{A} \,, \\
\label{A.prime}
\bar{\mathcal{A}}^\prime &= \mathrm{e}^{2\bar{\gamma}}\left( 2\bar{\gamma}^{\,\prime}\mathcal{A} + \bar{f}^{\prime}\mathcal{A}^\prime \right) \,, \\
\label{A.second}
\bar{\mathcal{A}}^{\prime\prime} &= \mathrm{e}^{2\bar{\gamma}}\left( 2\bar{\gamma}^{\,\prime\prime}\mathcal{A} + 4\left(\bar{\gamma}^{\,\prime}\right)^2\mathcal{A} + 4 \bar{\gamma}^{\,\prime} \bar{f}^\prime \mathcal{A}^\prime + \left(\bar{f}^\prime\right)^2\mathcal{A}^{\prime\prime} + \bar{f}^{\prime\prime}\mathcal{A}^\prime \right) \,.
\end{align}
From \eqref{A} we see that a diverging $\bar{\gamma}$ would render $\bar{\mathcal{A}}$ infinite because we have assumed $0 < \mathcal{A} < \infty$. Due to the latter without finetuning $\left|\bar{\gamma}^{\,\prime}\right| \to \infty$ implies $\left| \bar{\mathcal{A}}^{\prime} \right| \to \infty$ and analogously from \eqref{A.second} for the relation between $\bar{\gamma}^{\,\prime\prime}$ and $\bar{\mathcal{A}}^{\prime\prime}$. Hence, if we have two parametrizations where the restrictions \eqref{conditions.on.A.etc}-\eqref{conditions.on.A.second.etc} imposed on the three arbitrary functions $\left\lbrace \mathcal{A},\, \mathcal{V},\, \alpha \right\rbrace$ hold then the local Weyl rescaling connecting these parametrizations must be regular. In the spirit of the discussion around \eqref{conclusion.for.gamma.prime} let us consider the case $\bar{f}^{\prime} \to \infty$. From \eqref{A.prime} we obtain that in the same process $\mathcal{A}^\prime$ must vanish for the limiting value $\lim \bar{f}^\prime \mathcal{A}^\prime < \infty$ to hold because otherwise $\bar{\mathcal{A}}^{\prime}$ would necessarily diverge. Again let us make use of the backward transformations \eqref{conformal.transformation.backwards}-\eqref{field.redefinition.backwards} in order to write the transformation \eqref{A.prime} backwards
\begin{equation}
\label{A.prime.backwards}
0\overset{!}{=}\mathcal{A}^\prime =\mathrm{e}^{2\gamma}\left( 2\gamma^\prime \bar{\mathcal{A}} + f^\prime \bar{\mathcal{A}}^\prime \right) \,.
\end{equation}
Hence we conclude that in the process under consideration $f^{\prime} \to 0$ implies $\gamma^{\,\prime} \to 0$ and this is in a perfect agreement with the discussion after \eqref{conclusion.for.gamma.prime}.

Note that in the context of the regular Weyl rescaling \eqref{conformal.transformation} the conditions \eqref{conditions.on.A.etc}, i.e.~$0 < \mathcal{A} < \infty$ and $0 < \mathrm{e}^{2\alpha} < \infty$ are mathematical necessities for the existence of the transformations from an arbitrary frame to the Einstein frame \eqref{Einstein.frame} $(\mathcal{A}_\mathrm{E}=1)$ and to the Jordan frame \eqref{Jordan.frame} $(\alpha_\mathrm{J}=0)$, respectively.

%%%%%%%%%%%%%%%%%%%%%%%%%%%%%%%%%%%%%%%%%%%%%%%%%%%%%%%%%%%%%%%%%%%%%%%%%%%%%%%%%%%%%%%%%%%%%%%%%%%%%%%%%%%%%%%%%%%%%%%%%%%%%%%%%%%%%%%%

\subsubsection{Constraints on $\bar{f}$}

Let us recall that the function $\bar{f}(\bar{\Phi})$ is imposed to be at least directionally continuous. However it might be the case that $\bar{f}^\prime = 0$ or $\left| \bar{f}^\prime \right| \to \infty$. The latter has already been used implicitly because according to \eqref{definition.of.F} in EF canonical parametrization \eqref{Einstein.frame} $\bar{\mathcal{F}}_\mathrm{E} = 1$. Therefore if $\mathcal{F}$ diverges in some other parametrization then also $\bar{f}^{\prime}=0 \, \Leftarrow\, \bar{\mathcal{F}}_\mathrm{E} = \left( \bar{f}^\prime \right)^2\mathcal{F}$.

Let us continue analysing an analogous case more generically. We consider having $\mathcal{F} \to \infty$, $\bar{\mathcal{F}} < \infty$ and $\bar{f}^\prime = 0$. We proceed under the differentiable $\frac{1}{\mathcal{F}}$ assumption \eqref{non.diverging.derivatives.of.1.over.F}. Due to the transformation properties of $\mathcal{F}$ itself, i.e.\ \eqref{definition.of.F}, the transformation of the first derivative reads
\begin{equation}
\label{transformation.of.1.over.F.prime}
\left(\frac{1}{\mathcal{F}}\right)^\prime = \frac{2\bar{f}^{\prime\prime}}{\bar{\mathcal{F}}} + \bar{f}^\prime \left(\frac{1}{\bar{\mathcal{F}}}\right)^\prime \,.
\end{equation}
According to the nonvanishing $\left( \frac{1}{\mathcal{F}} \right)^\prime$ assumption \eqref{non.vanishing.first.derivative.of.1.over.F} the l.h.s.~of \eqref{transformation.of.1.over.F.prime} is nonzero. The second term on the r.h.s.~is zero and hence the first term on the r.h.s.~must be nonzero. The positive $\mathcal{F}$ assumption \eqref{nonvanishing.F} states that $\bar{\mathcal{F}}$ is nonvanishing and therefore also $\bar{f}^{\prime\prime} \neq 0$ in the case under consideration.

\renewcommand{\arraystretch}{2.5}
\begin{table}
	\fontsize{9.5}{8.5}\selectfont
	\caption{Conditions on transformation function $\bar{f}$ based on definition \eqref{definition.of.F}, assumptions \eqref{nonvanishing.F}-\eqref{non.vanishing.first.derivative.of.1.over.F} and rule \eqref{transformation.of.1.over.F.prime}. }
	\label{tab:1}
	\begin{tabular}{ccccccc}
		\cline{2-7}
		%first row
		\multirow{2}{*}{} & \multicolumn{2}{c}{$\bar{f}^\prime \rightarrow 0$} & \multicolumn{2}{c}{$0 < \left|\bar{f}^\prime\right| < \infty $} & \multicolumn{2}{c}{$\left|\bar{f}^\prime\right| \rightarrow \infty$}  \\ 
		\cline{2-7}
		%second row
		& $\bar{\mathcal{F}} < \infty $ & $\bar{\mathcal{F}} \rightarrow \infty $ & $\bar{\mathcal{F}} < \infty $ & $\bar{\mathcal{F}} \rightarrow \infty $ & $\bar{\mathcal{F}} < \infty $ & $\bar{\mathcal{F}} \rightarrow \infty $ \\
		\hline
		%third row
		\multirow{4}{*}{$\mathcal{F} < \infty $} & \multirow{4}{*}{-} & \multirow{4}{*}{-} & \multicolumn{1}{l}{iii) }  & \multirow{4}{*}{-} & \multirow{4}{*}{-} & \multicolumn{1}{l}{v) } \\[-0.2cm]
		%fourth row
		& & & $0 \leq \left|\bar{f}^{\prime\prime}\right| < \infty$ & & & $\left|\bar{f}^{\prime\prime}\right| \rightarrow \infty$	\\
		%fifth row
		& & & $0 \leq \displaystyle{\left|\frac{\bar{f}^{\prime\prime}}{\bar{\mathcal{F}}}\right|} < \infty$ & & & $0 <\displaystyle{\left| \frac{\bar{f}^{\prime\prime}}{\left(\bar{f}^\prime\right)^3} \right|} <\infty$ \\
		%sixth
		&&&&&& $0 < \displaystyle{\left| \frac{\bar{f}^{\prime\prime}}{\bar{f}^\prime\bar{\mathcal{F}}} \right|} < \infty$ \\
		\hline
		%first row of the lower part
		\multirow{6}{*}{$\mathcal{F} \rightarrow \infty $} &  
		\multicolumn{1}{l}{i) } & \multicolumn{1}{l}{ii) }  & \multirow{6}{*}{-} & \multicolumn{1}{l}{iv) a) } & \multirow{6}{*}{-} & \multicolumn{1}{l}{vi) } \\[-0.2cm]
		%second row of the lower part
		& \multirow{2}{*}{$0 < \left|\bar{f}^{\prime\prime}\right| < \infty$} & \multirow{2}{*}{$\left|\bar{f}^{\prime\prime}\right| \rightarrow \infty$}  & &  $ \left|\bar{f}^{\prime\prime}\right| < \infty $ & & \multirow{2}{*}{$\left|\bar{f}^{\prime\prime}\right| \rightarrow \infty$} \\
		%third row of the lower part
		& & & & $0 \leq \displaystyle{\left|\frac{\bar{f}^{\prime\prime}}{\bar{\mathcal{F}}}\right|} < \infty$&& \\
		%fourth row of the lower part
		\cline{5-5}
		& \multirow{3}{*}{$0 < \displaystyle{\left|\frac{\bar{f}^{\prime\prime}}{\bar{\mathcal{F}}}\right|} < \infty$} & \multirow{3}{*}{$0 < \displaystyle{\left|\frac{\bar{f}^{\prime\prime}}{\bar{\mathcal{F}}}\right|} < \infty$} & & \multicolumn{1}{l}{iv) b) }& & \multirow{3}{*}{$0 < \displaystyle{\left|\frac{\bar{f}^{\prime\prime}}{\bar{f}^\prime\bar{\mathcal{F}}}\right|} < \infty$} \\[-0.2cm]
		%fifth row of the lower part
		& & & & $\bar{f}^{\prime\prime} \to \infty $ & & \\
		%sixth row of the lower part
		& & & & $0 \leq \displaystyle{\left|\frac{ \bar{f}^{\prime\prime} }{ \bar{\mathcal{F}} }\right|} < \infty$ & & \\
		\hline
	\end{tabular}
\end{table} 
\renewcommand{\arraystretch}{1}

Table~\ref{tab:1} maps all possibilities for transformations between $\mathcal{F}$ and $\bar{\mathcal{F}}$. Here six situations can be considered, but not all of them are distinct. The two possibilities v) and vi) for which $\bar{f}^\prime \to \infty$ are taken into account by looking the two possibilities i) and ii) for $\bar{f}^\prime \to 0$ backwards. In order to examine the viability of the remaining four let us analyze each case separately.
\begin{itemize}
	\item[i)   ] The case $\bar{f}^{\prime} \to 0$ and $0 < \left| \bar{f}^{\prime\prime} \right| < \infty$.\\
	It does not have any pathologies so we keep it.
	
	\item[ii)  ] The case $\bar{f}^{\prime} \to 0$ and $ \left| \bar{f}^{\prime\prime} \right| \to \infty$. \\
	In order to reveal a pathology let us consider a transformation where JF BDBW quantities are considered to be the ``unbarred'' ones. Therefore $\mathcal{A}^\prime = 1$ and the transformation \eqref{A.second} implies $\bar{\mathcal{A}}^{\prime\prime} \to \infty$ which is something we want to avoid. Despite the fact that we used JF BDBW parametrization this behaviour is fairly general because of the assumption \eqref{nonvanishing.A.prime.etc} arising from first order small source conditions \eqref{condition.for.C.2}-\eqref{condition.for.C.3}. Due to such a pathology we neglect this possibility.
	
	\item[iii) ] The case $0< \left|\bar{f}^{\prime}\right| < \infty$ and $\left| \bar{f}^{\prime\prime} \right| < \infty$. \\
	This transformation is also perfectly normal and we keep it.
	
	\item[iv)  ]
	
	\begin{itemize}
		\item[a) ]  Almost the same as previous. Only that in this case both $\mathcal{F}$ and $\bar{\mathcal{F}}$ diverge. We keep it. 
		
		\item[b) ] The case $0< \left|\bar{f}^{\prime}\right| < \infty$ and $\left| \bar{f}^{\prime\prime} \right| \to \infty$. \\
		This case possess the same pathology as the case ii) and therefore we neglect it. 
	\end{itemize}

\end{itemize}
So, we will focus upon two possible cases. We shall refer to them according to the characteristics of the transformation function $\bar{f}$.
\begin{itemize}
	
	\item[a) ] `The regular case', based on the cases iii) and iv) a) in table~\ref{tab:1},
	\begin{equation}
	\label{regular.case}
	\begin{array}{c}
	0 < \left|\bar{f}^\prime\right| < \infty \,,\qquad \left| \bar{f}^{\prime\prime} \right| < \infty \,; \\
	% tühi rida
	\\
	\mathcal{F} < \infty \,,\quad \bar{\mathcal{F}} < \infty \qquad\text{or}\qquad \mathcal{F} \to \infty \,,\quad \bar{\mathcal{F}} \to \infty \,.
	\end{array} 
	\end{equation}
	
	\item[b) ] `The singular case', based on the case i) in table~\ref{tab:1},
	\begin{equation}
	\label{singular.case}
	\begin{array}{c}
	\bar{f}^\prime \to 0 \,,\qquad 0 < \left| \bar{f}^{\prime\prime} \right| < \infty \,,\qquad \bar{\gamma}^{\,\prime} \to 0 ; \\
	% empty row
	\\
	\mathcal{F} \to \infty \,,\qquad \bar{\mathcal{F}} < \infty \,.
	\end{array}
	\end{equation}
	
\end{itemize} 
We have also explicitly included the knowledge about $\bar{\gamma}^\prime$ given by discussion after \eqref{conclusion.for.gamma.prime} or equivalently after \eqref{A.prime.backwards}.

%%%%%%%%%%%%%%%%%%%%%%%%%%%%%%%%%%%%%%%%%%%%%%%%%%%%%%%%%%%%%%%%%%%%%%%%%%%%%%%%%%%%%%%%%%%%%%%%%%%%%%%%%%%%%%%%%%

\subsubsection{Transformation of the GR regime}\label{TOGRR}

In Subsec.~\ref{subsec:general.relativity.regime} we gave a notion of the GR regime and it is important to ascertain whether the given notion is invariant under the local rescaling of the metric tensor \eqref{conformal.transformation} and under the scalar field redefinition \eqref{field.redefinition}. Let us start by focusing upon the vanishing source conditions \eqref{potential.condition}-\eqref{matter.condition} for the GR regime at $\Phi_0$. Due to the constraints \eqref{conditions.on.A.etc}-\eqref{conditions.on.A.second.etc} imposed on the three arbitrary functions $\left\lbrace \mathcal{A},\, \mathcal{V},\, \alpha \right\rbrace$ and to the positive $\mathcal{F}$ assumption \eqref{nonvanishing.F} the following holds \cite{JKSV:2,JKSV:1}
\begin{align}
\label{potential.condition.invariant}
\frac{\mathcal{I}_2^\prime}{2\mathcal{I}_3^\prime} \equiv \pm \frac{ \mathcal{I}_2^\prime  }{2\sqrt{\mathcal{F}}} \equiv \pm \frac{ \left( \mathcal{A} \mathcal{V}^\prime - 2 \mathcal{V}\mathcal{A}^\prime \right) }{\mathcal{A}^2 \sqrt{ 2\mathcal{A}\mathcal{B} + 3 \left( \mathcal{A}^\prime \right)^2 } } = 0 \quad &\Leftrightarrow \quad \frac{\mathcal{A} \mathcal{I}_2^\prime}{4 \mathcal{F}} \equiv \frac{\left( \mathcal{A}\mathcal{V}^\prime  -  2 \mathcal{V} \mathcal{A}^\prime \right)}{2\mathcal{A}\mathcal{B}  +  3\left( \mathcal{A}^\prime\right)^2} = 0 \,, \\
\label{matter.condition.invariant}
\frac{ \left( \ln \mathcal{I}_1 \right)^\prime }{ 2 \mathcal{I}_3^\prime } \equiv \pm \frac{ \left( \ln \mathcal{I}_1 \right)^\prime }{2 \sqrt{\mathcal{F}}} \equiv \pm \frac{ \left( 2\mathcal{A}\alpha^\prime  -  \mathcal{A}^\prime \right)}{\sqrt{ 2\mathcal{A}\mathcal{B}  +  3\left( \mathcal{A}^\prime\right)^2} } = 0 \quad &\Leftrightarrow \quad \frac{ \left( \ln \mathcal{I}_1 \right)^\prime}{ 4\mathcal{A} \mathcal{F} } \equiv \frac{\left( 2\mathcal{A}\alpha^\prime  -  \mathcal{A}^\prime \right)}{2\mathcal{A}\mathcal{B}  +  3\left( \mathcal{A}^\prime\right)^2} = 0 \,.
\end{align}
The expressions on the left of both \eqref{potential.condition.invariant}-\eqref{matter.condition.invariant} are invariants in the spirit of Subsec.~\ref{Invariants} and therefore their numerical value does not depend on the parametrization. On the right of \eqref{potential.condition.invariant}-\eqref{matter.condition.invariant} are the vanishing source conditions \eqref{potential.condition}-\eqref{matter.condition} respectively. Hence we see that if these conditions hold in one parametrization then they hold in any other. In what follows the l.h.s.-s of \eqref{potential.condition.invariant}-\eqref{matter.condition.invariant} are referred to as the `invariant vanishing source conditions'.

The derivative of the scalar field $\Phi$ with respect to the spacetime coordinate transforms as follows: $\nabla_\mu \Phi = \bar{f}^\prime \bar{\nabla}_\mu \bar{\Phi}$. Hence for the regular case \eqref{regular.case} if $\nabla_\mu \Phi = 0$ then also $\bar{\nabla}_\mu \bar{\Phi} = 0$. For the singular case \eqref{singular.case} the latter does not have to be zero since $\bar{f}^\prime = 0$. Hence it might seem that the notion of the GR regime, i.e.~$\Phi=\Phi_0$, $\left.\nabla_\mu \Phi\right|_{\Phi_0} =0$, is not invariant. Nevertheless, taking into account the equation of motion \eqref{scalar.field.equation.without.R.divided.by.F} for the scalar field allows us to overcome this problem. Namely, let us proceed by considering the transformation
\begin{equation}
\label{transformation.of.second.derivative.of.Phi}
\nabla_\mu\nabla_\nu \Phi = \bar{f}^\prime \bar{\nabla}_\mu \bar{\nabla}_\nu \bar{\Phi} + \bar{f}^{\prime\prime} \bar{\nabla}_\mu \bar{\Phi} \bar{\nabla}_\nu \bar{\Phi} + \bar{f}^{\prime} \bar{\gamma}^{\,\prime} \left( \bar{g}_{\mu\nu} \bar{g}^{\sigma\rho} \bar{\nabla}_\sigma \bar{\Phi} \bar{\nabla}_\rho \bar{\Phi} - 2 \bar{\nabla}_\mu \bar{\Phi} \bar{\nabla}_\nu \bar{\Phi} \right) \,. 
\end{equation}
Again for the regular case \eqref{regular.case} if $\nabla_\mu \Phi = 0$ and $\nabla_\mu\nabla_\nu \Phi = 0$ then also $\bar{\nabla}_\mu \bar{\Phi}=0$ and $\bar{\nabla}_\mu \bar{\nabla}_\nu \bar{\Phi} = 0$. However in the singular case $\nabla_\mu\nabla_\nu \Phi = 0$ implies $\bar{\nabla}_\mu\bar{\Phi} = 0$ because $\bar{f}^{\prime\prime} \neq 0$. Therefore $\nabla_\mu \Phi = 0$ is preserved for both the regular \eqref{regular.case} and the singular \eqref{singular.case} transformation. Last but not least as all other terms in \eqref{scalar.field.equation.without.R.divided.by.F} are zero also $\bar{\rule{0ex}{1.5ex}\Box} \bar{\Phi}$ must vanish. The latter is automatically fulfilled if $\bar{\nabla}_\mu \bar{\nabla}_\nu \bar{\Phi}=0$ and we conclude that the GR regime is preserved under the local Weyl rescaling \eqref{conformal.transformation} and under the scalar field redefinition \eqref{field.redefinition}.

Let us point out that in the singular case \eqref{singular.case} the scalar field value $\Phi_\star$, defined by \eqref{star.critical.point}, gets transformed into $\Phi_\bullet$, given by \eqref{moon.critical.point}. However the vanishing source conditions \eqref{potential.condition}-\eqref{matter.condition} are fulfilled for both cases. Note also that due to the invariant vanishing source conditions \eqref{potential.condition.invariant}-\eqref{matter.condition.invariant}
\begin{equation}
\left\lbrace \Phi_\star \right\rbrace \cup \left\lbrace \Phi_\bullet \right\rbrace = \bar{f}\left( \left\lbrace \bar{\Phi}_\star \right\rbrace \cup \left\lbrace \bar{\Phi}_\bullet \right\rbrace \right) \,
\end{equation}
holds. In the following we will not distinguish between the elements of the same set.

%%%%%%%%%%%%%%%%%%%%%%%%%%%%%%%%%%%%%%%%%%%%%%%%%%%%%%%%%%%%%%%%%%%%%%%%%%%%%%%%%%%%%%%%%%%%%%%%%%%%%%%%%%%%%%%%%%

\subsubsection{Transformation of the hyperbolic critical points}

Let us consider the transformation of the first order small source conditions \eqref{condition.for.C.2}-\eqref{condition.for.C.3}. Here we do not provide a thorough analysis but rather give an insightful explanation. The condition \eqref{condition.for.C.2} is discussed in more details in Sec.~\ref{dynamical.system}.

We start by pointing out that the conditions under consideration are not given by invariants in the sense of Subsec.~\ref{Invariants} but they are closely related to the following invariant objects
\begin{align}
\label{nonhyperbolic.potential.condition.invariant}
\frac{1}{\mathcal{I}_3^\prime} \left( \frac{\mathcal{I}_2^\prime}{\mathcal{I}_3^\prime} \right)^\prime &\equiv \frac{1}{\pm\sqrt{\mathcal{F}}} \left( \frac{\mathcal{I}_2^\prime}{\pm\sqrt{\mathcal{F}}} \right)^\prime = \frac{\mathcal{I}_2^{\prime\prime}}{ \mathcal{F} } + \frac{1}{2} \left( \frac{1}{\mathcal{F}} \right)^\prime \mathcal{I}_2^\prime \,,\\
\label{nonhyperbolic.matter.condition.invariant}
\frac{1}{\mathcal{I}_3^\prime} \left( \frac{ \left( \ln \mathcal{I}_1 \right)^\prime }{ \mathcal{I}_3^\prime } \right)^\prime &\equiv \frac{1}{\pm\mathcal{F}} \left( \frac{ \left( \ln \mathcal{I}_1 \right)^\prime }{ \pm\sqrt{\mathcal{F}} } \right)^\prime = \frac{ \left( \ln \mathcal{I}_1 \right)^{\prime\prime}}{ \mathcal{F} } + \frac{1}{2} \left( \frac{1}{\mathcal{F}} \right)^\prime \left( \ln \mathcal{I}_1 \right)^\prime \,.
\end{align}
In principle we have taken the derivative of the invariant vanishing source conditions \eqref{potential.condition.invariant}-\eqref{matter.condition.invariant} w.r.t.~the invariant $\mathcal{I}_3$ \cite{JKSV:2} and in the following we shall refer to \eqref{nonhyperbolic.potential.condition.invariant}-\eqref{nonhyperbolic.matter.condition.invariant} as the `invariant first order small source conditions'. The quantities \eqref{nonhyperbolic.potential.condition.invariant}-\eqref{nonhyperbolic.matter.condition.invariant} in principle differ from the (noninvariant) first order small source conditions \eqref{condition.for.C.2}-\eqref{condition.for.C.3} only by a factor $\frac{1}{2}$ in front of the second additive term on the r.h.s. As mentioned in the discussion after \eqref{condition.for.C.3} one of the additive terms must be zero in the GR regime. The same holds for \eqref{nonhyperbolic.potential.condition.invariant}-\eqref{nonhyperbolic.matter.condition.invariant}. Hence we conclude that \eqref{condition.for.C.2}-\eqref{condition.for.C.3} are nonvanishing (vanishing) if and only if the invariants \eqref{nonhyperbolic.potential.condition.invariant}-\eqref{nonhyperbolic.matter.condition.invariant} respectively are nonvanishing (vanishing). In other words if the necessary conditions \eqref{condition.for.C.2}-\eqref{condition.for.C.3} are fulfilled in one parametrization then they are also satisfied in any other.

Therefore from \eqref{Einstein.frame.first.ord.small}-\eqref{Jordan.frame.first.ord.small} we obtain the following: if in JF BDBW parametrization \eqref{Jordan.frame} $\left( \frac{1}{2\omega + 3} \right)^\prime \neq 0$ then also for the same theory written in EF canonical parametrization \eqref{Einstein.frame} $\alpha^{\prime\prime} \neq 0$.

%%%%%%%%%%%%%%%%%%%%%%%%%%%%%%%%%%%%%%%%%%%%%%%%%%%%%%%%%%%%%%%%%%%%%%%%%%%%%%%%%%%%%%%%%%%%%%%%%%%%%%%%%%%%%%%%%%%%%%%%%%%%%%%%%%%%%%%%%%%%%%%%%%%%%%%%%%%%%%%%%%%%%%%%%%%%%%%%%%%%%%%%%%%%%%%%%%%%%%%%%%%%%%%%%%%%%%%%%%%%%%%%%%%%%%%%%%%%%%%%%%%%%%%%%%%%%%%%

\section{Dynamical system in the Friedmann cosmology}\label{dynamical.system}

The aim of this section is to work through a relatively simple example in the framework of the Friedmann cosmology (see Subsubsec.~\ref{Friedmann.cosmology}) in order to prove the following. Let us consider the GR regime as a hyperbolic critical point in the context of the dynamical systems approach. The qualitative behaviour of the critical point is determined by invariants and therefore whether the theory under consideration converges to GR or repels from it does not depend on the chosen parametrization. In order to show the nontriviality of the transformations we provide a lot of calculational details.

%%%%%%%%%%%%%%%%%%%%%%%%%%%%%%%%%%%%%%%%%%%%%%%%%%%%%%%%%%%%%%%%%%%%%%%%%%%%%%%%%%%%%%%%%%%%%%%%%%%%%%%%%%%%%%

\subsection{Theory: part III}

Let us start by using the notation of the current paper for rewriting the approach formulated in \cite{JKS2010,JKS:2010:time}. Our focus is upon the transformation properties. Therefore, in order to make our calculations less lengthy and more transparent we truncate the physical side of the theory by considering the flat Friedmann cosmology without matter, i.e.~$k=0 \text{ and } T_{\mu\nu} = 0$. Note that this entails the dropping of the coupling function $\alpha$ from the theory. Due to the latter the notion of the GR regime slightly differs from the one used in \cite{JKS2010} because one of the vanishing source conditions, namely \eqref{matter.condition} is no longer needed.

%%%%%%%%%%%%%%%%%%%%%%%%%%%%%%%%%%%%%%%%%%%%%%%%%%%%%%%%%%%%%%%%%%%%%%%%%%%%%%%%%%%%%%%%%%%%%%%%%%%%%%%%%%%%%%

\subsubsection{Critical points for potential $\mathcal{V}$ dominated universe with $\rho=0 \,, k=0 $}

We want to study the scalar field equation \eqref{Scalar.field.equation.in.FLRW.cosmology} as a dynamical system in order to write out the critical points and study their properties. Let us follow the well known prescription: solve the Friedmann constraint equation \eqref{First.Friedmann.general.equation} as a quadratic equation for $H$ and plug the answer into the scalar field equation of motion \eqref{Scalar.field.equation.in.FLRW.cosmology}. The resulting equation reads
\begin{equation}
\label{Dynamical.system}
\ddot{\Phi} = \frac{1}{2}\frac{\mathcal{A}^\prime}{\mathcal{A}} \dot{\Phi}^2 + \frac{1}{2}\left( \frac{1}{\mathcal{F}} \right)^\prime\mathcal{F} \dot{\Phi}^2 -  \dot{\Phi} \varepsilon \sqrt{3\mathcal{F}\dot{\Phi}^2 + 3\ell^{-2}\mathcal{I}_2\mathcal{A}} - \frac{\mathcal{A}}{2\ell^2} \frac{ \mathcal{I}_2^\prime }{ \mathcal{F} }  \,
\end{equation}
where $\varepsilon = +1$ ($\varepsilon=-1$) corresponds to the positive (negative) solution of \eqref{First.Friedmann.general.equation} as a quadratic equation for $H$, i.e.~in principle to the expanding (contracting) universe. Analogically to \eqref{scalar.field.equation.without.R.divided.by.F} we made use of the invariants, defined by \eqref{invariants}, and $\mathcal{F}$, given by \eqref{definition.of.F}, in order to write \eqref{Dynamical.system} in a more compact form. For a critical point one must impose $\dot{\Phi} = 0$. For the latter to be sustained also $\ddot{\Phi} = 0$ must hold, i.e.~the r.h.s.~of \eqref{Dynamical.system} must vanish. Hence the critical point corresponds to the GR regime discussed in Sec.~\ref{general.relativity.regime}. In the context of the latter the scalar field equation \eqref{Dynamical.system} describes a process that may approach that regime. In the current case we have omitted the influence of $\alpha$ and therefore the condition \eqref{matter.condition} or the equivalent invariant condition \eqref{matter.condition.invariant} is no longer needed. Hence the GR regime ($\nabla_\mu\Phi=0$ and the vanishing source conditions \eqref{potential.condition}-\eqref{matter.condition}) reduce to
\begin{equation}
\label{general.relativity.conditions.for.FLRW.cosmology}
\dot{\Phi} = 0 \,,\qquad \frac{ \mathcal{I}_2^\prime }{ \mathcal{F} } = 0 \,.
\end{equation}
Therefore, analogically to \eqref{moon.critical.point}-\eqref{star.critical.point} while also taking into account the first order small source condition \eqref{condition.for.C.2} we distinguish the scalar field values as
\begin{align}
\label{moon.critical.point.in.FLRW}
\Phi_\bullet &:\quad \left. \mathcal{I}_2^\prime \right|_{\Phi_\bullet} = 0 \quad \text{ and } \quad \left. \mathcal{I}_2^{\prime\prime} \right|_{\Phi_\bullet} \neq 0 \quad \text{ and }\quad \left. \frac{1}{\mathcal{F}} \right|_{\Phi_\bullet} \neq 0 \,, \\
\label{star.critical.point.in.FLRW}
\Phi_\star &:\quad \left. \frac{1}{\mathcal{F}} \right|_{\Phi_\star} = 0 \quad \text{ and } \quad  \left. \left(\frac{1}{\mathcal{F}}\right)^\prime \right|_{\Phi_\star} \neq 0 \quad \text{ and } \quad \left. \mathcal{I}_2^\prime \right|_{\Phi_\star} \neq 0 \,.
\end{align}
It is well known that for the scalar field value $\Phi_\bullet$ the equation \eqref{Dynamical.system} can be rewritten as an ordinary dynamical system \cite{Faraoni:Jensen:Theuerkauf,Leon,JKS2008} but the case corresponding to $\Phi_\star$ must be studied more carefully. Namely, if $\mathcal{F}$ diverges then \eqref{general.relativity.conditions.for.FLRW.cosmology} gives necessary but not sufficient conditions. From \eqref{Dynamical.system} one can see that $\left. \ddot{\Phi} \right|_{\substack{\Phi = \Phi_\star \\ \dot{\Phi} =0}} = 0$ can only be achieved if for a trajectory of a specific solution under consideration the limiting value
\begin{equation}
\label{limiting.value}
\lim\limits_{\substack{
		\Phi \to \Phi_\star \\
		\dot{\Phi} \rightarrow 0
	}
}\left\lbrace\left(\frac{1}{\mathcal{F}}\right)^\prime \mathcal{F}\dot{\Phi}^2 \right\rbrace= 0
\end{equation}
holds. In the case when the latter is violated a trajectory passes the point in the phase space where the conditions given by \eqref{general.relativity.conditions.for.FLRW.cosmology} are fulfilled, i.e.~the critical point. Hence e.g.~for an attractive critical point the limiting value \eqref{limiting.value} restricts the ``final'' part of the trajectory, i.e.~the order of the magnitude of $\dot{\Phi}$ relative to $\Phi-\Phi_\star$ in process where the scalar field $\Phi$ evolves toward $\Phi_\star$ and stops there. Taking into account the knowledge of \eqref{diverging.F.equivalent.to.one.over.x} we obtain that the expression under the limiting value \eqref{limiting.value} is equivalent to $\frac{\dot{x}^2}{x}$. Therefore it necessarily holds up to $\dot{\Phi}$ being the same or higher order small compared to $\Phi - \Phi_\star$ \cite{JKS2008}.

%%%%%%%%%%%%%%%%%%%%%%%%%%%%%%%%%%%%%%%%%%%%%%%%%%%%%%%%%%%%%%%%%%%%%%%%%%%%%%%%%%%%%%%%%%%%%%%%%%%%%%%%%%%%%%%%

\subsubsection{Perturbed equation}

Let us introduce the following notation for small perturbations
\begin{align}
\label{x}
x &\equiv \Phi - \Phi_0 \,, \\
\label{x.dot}
\dot{x} &\equiv \dot{\Phi} \,
\end{align}
where $\Phi_0$ is defined by either \eqref{moon.critical.point.in.FLRW} or by \eqref{star.critical.point.in.FLRW}.

Let us first write out the first order perturbed approximation of \eqref{Dynamical.system} as follows
\begin{equation}
\label{First.order.equation}
E^{(x)} \equiv - \ddot{x} +  \frac{M}{2}\frac{\dot{x}^2}{x} - C_1^{\varepsilon} \cdot \dot{x} + C_2 \cdot x = 0 \,.
\end{equation}
While deriving the first order perturbed equation \eqref{First.order.equation} for \eqref{Dynamical.system} we dropped the first term on the r.h.s.~of \eqref{Dynamical.system} because it is definitely higher order term. For the first order approximation of the third and the fourth term on the r.h.s.~of \eqref{Dynamical.system} we use the Taylor expansion and the following constants
\begin{equation}
\label{C.1.and.C.2}
C_1^{\varepsilon} \equiv \left. \frac{\varepsilon}{\ell} \sqrt{3 \mathcal{I}_2 \mathcal{A} } \right|_{x = 0} \,; \qquad C_2 \equiv - \left. \left( \frac{\mathcal{A} }{2\ell^2} \frac{ \mathcal{I}_2^\prime }{ \mathcal{F} } \right)^\prime \right|_{x = 0} \equiv - \left.\frac{ \mathcal{A} }{ 2\ell^2 }\right|_{x=0}  \lim\limits_{x \to 0} \left[  ^{\displaystyle{ \frac{ \mathcal{I}_2^\prime }{ \mathcal{F} } } } \Big/ _{\displaystyle{x}} \right]  \,.
\end{equation}
The last equality for $C_2$ holds because if $x\to 0$ then due to the critical point condition \eqref{general.relativity.conditions.for.FLRW.cosmology} the numerator vanishes and we can make use of the l'Hospital rule. The constant $C_2$ is the same as defined by the first order small source condition \eqref{condition.for.C.2}.

The second term on the r.h.s.~of \eqref{Dynamical.system} is a bit tricky. Namely if $\mathcal{F}$ is finite then this is already higher order term and we drop it but if $\mathcal{F}$ diverges then calculating the Taylor expansion introduces coefficients that depend on the ratio $\frac{\dot{x}}{x}$. The latter is clearly something we want to avoid because the properties of a critical point should not depend on the choice of the trajectory. We instead make use of the knowledge obtained by the second remark \eqref{diverging.F.equivalent.to.one.over.x} in Subsubsec.~\ref{Two.remarks}. Hence, due to the nonvanishing $\left( \frac{1}{\mathcal{F}} \right)^\prime$ assumption \eqref{non.vanishing.first.derivative.of.1.over.F}, if $\mathcal{F}$ diverges then
\begin{equation}
\label{substitution}
\frac{1}{2}\left( \frac{1}{\mathcal{F}} \right)^\prime\mathcal{F} \dot{\Phi}^2 \sim \frac{1}{2}\frac{\dot{x}^2}{x}
\end{equation}
holds. In the diverging $\mathcal{F}$ case this is the first order approximation to the limiting value \eqref{limiting.value}. In order to capture these two possibilities ($\mathcal{F}<\infty$ and $\mathcal{F}\to \infty$) into one equation we substitute the second term on the r.h.s.~of the scalar field equation \eqref{Dynamical.system} by
\begin{equation}
\label{M}
\frac{M}{2} \frac{\dot{x}^2}{x} \,,\qquad M \equiv \lim\limits_{x \to 0}\left\lbrace \left(\frac{1}{\mathcal{F}}\right)^\prime \mathcal{F}\cdot x\right\rbrace \,
\end{equation}
where $M$, introduced in \eqref{diverging.F.equivalent.to.one.over.x}, is a constant with the following properties. If $\mathcal{F}<\infty$ then $M=0$ and the term vanishes but if $\mathcal{F} \to \infty$ then $M=1$ and the term survives. Hence, we conclude that \eqref{First.order.equation} is the first order approximation of \eqref{Dynamical.system}.

%%%%%%%%%%%%%%%%%%%%%%%%%%%%%%%%%%%%%%%%%%%%%%%%%%%%%%%%%%%%%%%%%%%%%%%%%%%%%%%%%%%%%%%%%%%%%%%%%%%%%%%%%%%%%%%%

\subsubsection{Linearized equation}

We have obtained a first order approximated equation \eqref{First.order.equation} but in the case of the $\Phi_{\star}$ critical point \eqref{star.critical.point.in.FLRW}, where $\mathcal{F}$ diverges, this is a nonlinear equation and we cannot apply the usual methods of the dynamical systems directly. However it turns out that \eqref{First.order.equation} contains a hidden linearity. Let us make use of the coordinate transformation that was proposed in \cite{JKS2010}
\begin{equation}
\label{x.tilde}
\tilde{x} \equiv \pm \frac{x}{\left| x \right|^{\frac{M}{2}}} \,
\end{equation}
where the meaning of $\pm$ becomes clear later. The derivatives of $\tilde{x}$ with respect to cosmological time $t$ read
\begin{equation}
\label{x.tilde.dot}
\dot{\tilde{x}} = \pm \frac{\dot{x}}{\left| x \right|^{\frac{M}{2}}}\left( \frac{2-M}{2} \right) \,,\qquad \ddot{\tilde{x}} = \pm \left( \ddot{x} -\frac{M}{2} \frac{\dot{x}^2}{x} \right) \frac{1}{\left| x \right|^{\frac{M}{2}}} \frac{2-M}{2} \,.
\end{equation}
The obtained results can be used to rewrite \eqref{First.order.equation} as
\begin{equation}
\label{linear.perturbed.equation}
\mp \frac{2}{2-M} \left| x \right|^{\frac{M}{2}} \left\lbrace \ddot{\tilde{x}} + C_1^{\varepsilon} \dot{\tilde{x}} - \frac{2-M}{2}C_2\tilde{x} \right\rbrace = 0 \,.
\end{equation}
We are not interested in the trivial solution $x \equiv 0$. Therefore the expression in the curly brackets must be equal to zero and this is a linear equation which can be written as an ordinary dynamical system
\begin{equation}
\label{Dynamical.system.matrix.equation}
\begin{pmatrix}
	\dot{\tilde{x}} \\
	\dot{\tilde{y}}
\end{pmatrix}
=
\begin{pmatrix}
	0 & 1 \\
	\frac{2-M}{2}C_2 & -C_1^{\varepsilon}
\end{pmatrix}
\begin{pmatrix}
	\tilde{x} \\
	\tilde{y}
\end{pmatrix}
\,
\end{equation}
where $\tilde{y} \equiv \dot{\tilde{x}}$. In the following we shall use `linearized equation' to refer to the dynamical system \eqref{Dynamical.system.matrix.equation} or equivalently to the underlying expression in the curly brackets of \eqref{linear.perturbed.equation}. The term `linear equation' is used to denote the perturbed equation \eqref{First.order.equation} in the case when $M=0$. Note that for the latter the coordinate transformation \eqref{x.tilde} is actually an identity transformation up to a sign. Hence for the case $M=0$ the linear equation and the linearized equation coincide. The solutions and hence also the properties of the critical point are now determined by the eigenvalues of the matrix that contains the constant coefficients.

%%%%%%%%%%%%%%%%%%%%%%%%%%%%%%%%%%%%%%%%%%%%%%%%%%%%%%%%%%%%%%%%%%%%%%%%%%%%%%%%%%%%%%%%%%%%%%%%%%%%%%%%%%%%%%%%

\subsubsection{Solutions}

The eigenvalues of the square matrix in \eqref{Dynamical.system.matrix.equation} are
\begin{equation}
\label{eigenvalue}
\lambda^{\varepsilon}_{\pm} = \frac{1}{2}\left( -C_1^{\varepsilon} \pm \sqrt{ \left( C_1^{\varepsilon} \right)^2 + 2(2-M)C_2 } \right) \,.
\end{equation}
The latter can be used to write out the solution for $\tilde{x}$. In order to determine the behaviour of $x \equiv \Phi - \Phi_0$ we invert the relation \eqref{x.tilde} as
\begin{equation}
x = \pm \left| \tilde{x} \right|^{\frac{2}{2-M}} \,.
\end{equation}
If the eigenvalues $\lambda_+$ and $\lambda_-$ are different then the solution for $x$ reads
\begin{equation}
\label{solution.for.x}
x(t) = \pm \left( K_1 \mathrm{e}^{\lambda^{\varepsilon}_{+} t} + K_2 \mathrm{e}^{ \lambda^{\varepsilon}_{-} t} \right)^{\frac{2}{2-M}}  \,
\end{equation}
where $K_1$ and $K_2$ are integration constants. In principle the same result was obtained in \cite{JKSV:2}. Due to the power $\frac{2}{2-M}$ the theory under consideration is indeed endowed with a mechanism called for in the first remark of Subsubsec.~\ref{Two.remarks}. Namely, the diverging $\mathcal{F}$ implies $M=1$ as mentioned in the discussion after \eqref{M} and in that case $x = \pm \tilde{x}^2$. Therefore we have an encoded mechanism which due to the nonvanishing $\left( \frac{1}{\mathcal{F}} \right)^\prime$ assumption \eqref{non.vanishing.first.derivative.of.1.over.F} does not allow to violate the positive $\mathcal{F}$ assumption \eqref{nonvanishing.F}. In the case $M=0$ the power $\frac{2}{2-M} = 1$ because, as mentioned earlier, then the coordinate transformation \eqref{x.tilde} is an identity transformation up to a sign.

The $\pm$ in coordinate transformation \eqref{x.tilde} can now be reasoned as follows. If $M=0$ then it does not matter whether $x = \tilde{x}$ or $x = -\tilde{x}$ but if $M=1$ then in the spirit of the first remark in Subsubsec.~\ref{Two.remarks} one must have $x \geq 0$ $\left( x \leq 0 \right)$ if $\left(\frac{1}{\mathcal{F}} \right)^\prime > 0$  $\left( \left(\frac{1}{\mathcal{F}} \right)^\prime < 0 \right)$. The same applies to the solution \eqref{solution.for.x} and there the sign `$+$' (`$-$') corresponds to $\left(\frac{1}{\mathcal{F}} \right)^\prime > 0$ $\left( \left(\frac{1}{\mathcal{F}} \right)^\prime < 0 \right)$. Because of the previous reasoning we have also dropped the absolute value in \eqref{solution.for.x}.

Let us point out that in case of the diverging $\mathcal{F}$ if the nonvanishing $\left( \frac{1}{\mathcal{F}} \right)^\prime$ assumption \eqref{non.vanishing.first.derivative.of.1.over.F} is not fulfilled then $C_2=0$ as can be read out from \eqref{C.1.and.C.2} or equivalently from \eqref{condition.for.C.2}. Therefore one eigenvalue is zero and the critical point is nonhyperbolic. We conclude that the first order small source conditions \eqref{condition.for.C.2}-\eqref{condition.for.C.3} in Subsubsec.~\ref{subsec:Dynamical.approach.to.the.GR.regime} are indeed well motivated.

%%%%%%%%%%%%%%%%%%%%%%%%%%%%%%%%%%%%%%%%%%%%%%%%%%%%%%%%%%%%%%%%%%%%%%%%%%%%%%%%%%%%%%%%%%%%%%%%%%%%%%%%%%%%%%%%%%%%%%%%%%%%%%%%%%%%%%%%%%%%%%%%%%%%%%%%%%%%%%%%%%%%%%%%%%%%%%%%%%%%%%%%%%%%%%%%%%%%%%%%%%%%%%%%%%%%%%%%%%%%%%%%

\subsection{Transformations: part III}

\subsubsection{Transformation of the perturbed equation}

In order to study the transformation of the first order perturbed equation \eqref{First.order.equation}, let us first consider the transformation of $x$, $\dot{x}$ and $\ddot{x}$. For the latter we take the definitions \eqref{x}-\eqref{x.dot}, write these in terms of the ``barred'' quantities, i.e.\ make use of \eqref{field.redefinition}, \eqref{transformation.of.Phi.dot}, also taking into account the cosmological time transformation \eqref{time.coordinate.transformation}, and then use the Taylor expansion around $\bar{x}=0$, $\dot{\bar{x}}=0$ and $\ddot{\bar{x}}=0$ in order to obtain polynomials with respect to $\bar{x}$, $\dot{\bar{x}}$ and $\ddot{\bar{x}}$. At the moment we shall keep the terms up to the second order for reason that will become clear soon:
\begin{align}
\label{transformation.of.x}
x &\equiv \Phi - \Phi_0 = \bar{f}(\bar{\Phi}) - \bar{f}(\bar{\Phi})|_0 \approx \left[ \bar{f}^\prime \,\right]_0\bar{x} +  \frac{1}{2} \left[ \bar{f}^{\prime\prime} \,\right]_0\bar{x}^2 \,,\\
\label{transformation.of.x.dot}
\dot{x}&\equiv \frac{\mathrm{d}\phantom{t}}{\mathrm{d}t}\left( \Phi - \Phi_0 \right) = \mathrm{e}^{-\bar{\gamma}}\bar{f}^{\prime} \frac{\mathrm{d}\bar{\Phi}}{\mathrm{d}\bar{t}} \approx \left[ \mathrm{e}^{-\bar{\gamma}}\bar{f}^\prime \,\right]_0 \dot{\bar{x}} + \left[ \mathrm{e}^{-\bar{\gamma}}\bar{f}^{\prime\prime} \,\right]_0 \bar{x}\dot{\bar{x}} - \left[ \bar{\gamma}^\prime \mathrm{e}^{-\bar{\gamma}}\bar{f}^\prime \,\right]_0 \bar{x}\dot{\bar{x}} \,,\\
\nonumber
\ddot{x} &\equiv \frac{\mathrm{d}\phantom{t}}{\mathrm{d}t}\frac{\mathrm{d}\phantom{t}}{\mathrm{d}t} \left( \Phi - \Phi_0 \right) = \mathrm{e}^{-2\bar{\gamma}}\bar{f}^{\prime} \ddot{\bar{x}} + \mathrm{e}^{-2\bar{\gamma}}\bar{f}^{\prime\prime} \dot{\bar{x}}^2 - \mathrm{e}^{-2\bar{\gamma}} \bar{\gamma}^{\,\prime}\bar{f}^\prime \dot{\bar{x}}^2 \\
&\approx \left[ \mathrm{e}^{-2\bar{\gamma}}\bar{f}^{\prime} \,\right]_0 \ddot{\bar{x}} + \left[ \mathrm{e}^{-2\bar{\gamma}} \bar{f}^{\prime\prime} \,\right]_0 \dot{\bar{x}}^2 - \left[ \mathrm{e}^{-2\bar{\gamma}}\bar{\gamma}^{\,\prime} \bar{f}^{\prime} \,\right]_0 \dot{\bar{x}}^2 + \left[ \mathrm{e}^{-2\bar{\gamma}} \bar{f}^{\prime\prime} \,\right]_0 \bar{x} \ddot{\bar{x}} - 2\left[ \mathrm{e}^{-2\bar{\gamma}} \bar{\gamma}^{\,\prime} \bar{f}^\prime \,\right]_0 \bar{x} \ddot{\bar{x}} \,
\end{align}
where $\bar{\Phi}_0: \Phi_0 = \bar{f}(\bar{\Phi}_0)$. When calculating the transformation of the first order perturbed equation \eqref{First.order.equation} we keep only the leading order terms. Therefore for the regular case \eqref{regular.case} we substitute as follows:
\begin{equation}
\label{regular:x.etc.}
x = \left[ \bar{f}^\prime \,\right]_0 \bar{x} \,, \qquad \dot{x} = \left[ \mathrm{e}^{-\bar{\gamma}} \bar{f}^\prime \,\right]_0 \dot{\bar{x}} \,, \qquad \frac{\dot{x}^2}{x} = \left[ \mathrm{e}^{-2\bar{\gamma}} \bar{f}^\prime \,\right]_0 \frac{\dot{\bar{x}}^2}{\bar{x}}  \,,\qquad \ddot{x} = \left[ \mathrm{e}^{-2\bar{\gamma}} \bar{f}^\prime \,\right]_0 \ddot{\bar{x}} \,.
\end{equation}
However in the singular case \eqref{singular.case} the linear order coefficients vanish due to $\left. \bar{f}^\prime \right|_0 = 0$ and the leading order is actually quadratic. Hence, in the singular case \eqref{singular.case} we substitute as
\begin{align}
\nonumber
x &= \frac{1}{2} \left[ \bar{f}^{\prime\prime} \,\right]_0 \bar{x} \bar{x} \,, \qquad \dot{x} = \left[ \mathrm{e}^{-\bar{\gamma}} \bar{f}^{\prime\prime} \,\right]_0 \bar{x} \dot{\bar{x}} \,, \qquad \frac{\dot{x}^2}{x} =  2 \left[ \mathrm{e}^{-2\bar{\gamma}} \bar{f}^{\prime\prime} \,\right]_0 \bar{x} \frac{\dot{\bar{x}}^2}{\bar{x}}  \,, \\
\label{singular:x.etc.}
\ddot{x} &= \left[ \mathrm{e}^{-2\bar{\gamma}} \bar{f}^{\prime\prime} \,\right]_0 \bar{x} \ddot{\bar{x}} +  \left[ \mathrm{e}^{-2\bar{\gamma}} \bar{f}^{\prime\prime} \,\right]_0 \bar{x} \frac{\dot{\bar{x}}^2}{\bar{x}} \,.
\end{align}
Let us point out that in case of the singular transformation \eqref{singular.case} the order of magnitude of the small perturbation $\Phi - \Phi_0$ changes, i.e.~$x$ that is first order small in its own parametrization is actually second order small with respect to $\bar{x}$. Note also that $\dot{x}=0$ whenever $\bar{x}=0$.

One can show that the coefficients $M\,,C_1^{\varepsilon}\,,C_2$, defined by \eqref{M} and \eqref{C.1.and.C.2} respectively, transform as follows
\begin{align}
\label{Mteisenemine}
M &= \bar{Q}_1\bar{M} + \bar{Q}_2  \,, \\
\label{C.1.teisenemine}
C_1^{\varepsilon} &= \left.\mathrm{e}^{-\bar{\gamma}}\right|_{\bar{\Phi}_0}\bar{C}_1^{ \varepsilon } \,, \\
\label{C.2.teisenemine}
C_2 &= \left.\mathrm{e}^{-2\bar{\gamma}}\right|_{\bar{\Phi}_0}\bar{Q}_1^{-1}\bar{C}_2 \,
\end{align}
where $\bar{Q}_1$ and $\bar{Q}_2$ are the limiting values
\begin{align}
\label{Q.1}
\bar{Q}_1 &\equiv \lim\limits_{\bar{\Phi} \to \bar{\Phi}_0}\left\lbrace \frac{\bar{f}(\bar{\Phi}) - \bar{f}(\bar{\Phi})|_0}{\bar{f}^\prime \cdot \left(\bar{\Phi} - \bar{\Phi}_0\right)} \right\rbrace \,,
\\
\label{Q.2}
\bar{Q}_2 &\equiv \lim\limits_{\bar{\Phi}\to \bar{\Phi}_0}\left\lbrace \frac{2\bar{f}^{\prime\prime}\left(\bar{f}(\bar{\Phi})- \bar{f}(\bar{\Phi})|_0\right)}{\left(\bar{f}^\prime\right)^2} \right\rbrace \,.
\end{align}
The assumption that $\bar{f}(\bar{\Phi})$ itself is at least directionally continuous implies that if $\Phi \to \Phi_0$ $\left( x \to 0 \right)$ then also $\bar{\Phi} \to \bar{\Phi}_0$ $\left( \bar{x} \to 0 \right)$. Hence when calculating the limiting values we actually do not have to concern about the limiting process itself. The limiting values \eqref{Q.1}-\eqref{Q.2} can be calculated for example by making use of the l'Hospital rule. The results read
\begin{align}
\label{Q.1.values}
\bar{Q}_1 &= \left\lbrace  
\begin{array}{cl}
1&\text{ for the regular case \eqref{regular.case}} \\
\frac{1}{2}&\text{ for the singular case \eqref{singular.case}}
\end{array}
\right. \,, \\
\label{Q.2.values}
\bar{Q}_2 &= \left\lbrace  
\begin{array}{cl}
 	0 &\text{ for the regular case \eqref{regular.case}} \\
	1 &\text{ for the singular case \eqref{singular.case}}
\end{array}
\right. \,.
\end{align}
Let us point out that based on the definition of $M$, given by \eqref{M}, one can determine from table~\ref{tab:1} that
\begin{equation}
\label{M.values}
\begin{array}{cl}
M = 0 \text{  and  } \bar{M}=0 \qquad \text{or} \qquad M = 1 \text{  and  }\bar{M} = 1 \qquad &\text{for the regular case \eqref{regular.case}} \,,\\
M = 1 \text{  and  } \bar{M} = 0
\qquad &\text{for the singular case \eqref{singular.case}} \,.
\end{array}
\end{equation}
This result is consistent with the transformation rule \eqref{Mteisenemine} for $M$ and the results \eqref{Q.1.values}-\eqref{Q.2.values}.

By making use of the transformation rules \eqref{regular:x.etc.}, \eqref{Mteisenemine}-\eqref{C.2.teisenemine} and results \eqref{Q.1.values}-\eqref{Q.2.values} we obtain that in the regular case the transformation of the perturbed equation \eqref{First.order.equation} reads
\begin{align}
\nonumber E^{(x)} &= -\ddot{x} + \frac{M}{2}\frac{\dot{x}^2}{x} - C_1^{\varepsilon} \dot{x} + C_2 x = \\
\label{transformation.of.perturbed.equation.regular}
&= \left[ \mathrm{e}^{-2\bar{\gamma}}\bar{f}^\prime \,\right]_0 \left\lbrace -\ddot{\bar{x}} + \frac{1}{2}\left( 1 \cdot \bar{M} + 0 \right) \frac{\dot{\bar{x}}^2}{\bar{x}} - \bar{C}_1^{\varepsilon} \dot{\bar{x}} + 1 \cdot \bar{C}_2 \bar{x} \right\rbrace = \left[ \mathrm{e}^{-2\bar{\gamma}}\bar{f}^\prime \,\right]_0 \bar{E}^{(\bar{x})} \,.
\end{align}
Analogously by using \eqref{singular:x.etc.} etc.~for the singular case \eqref{singular.case} the transformation reads
\begin{equation}
E^{(x)} = \left[ \mathrm{e}^{-2\bar{\gamma}}\bar{f}^{\prime\prime} \,\right]_0 \bar{x} \left\lbrace -\ddot{\bar{x}} + \frac{2}{2}\left( \frac{1}{2} \negmedspace \cdot \negmedspace \bar{M} \negmedspace + \negmedspace 1 \negmedspace - \negmedspace 1 \right) \frac{\dot{\bar{x}}^2}{\bar{x}} - \bar{C}_1^{\varepsilon} \dot{\bar{x}} + \frac{1}{2} \negmedspace \cdot \negmedspace 2 \bar{C}_2 \bar{x} \right\rbrace = \left[ \mathrm{e}^{-2\bar{\gamma}}\bar{f}^{\prime\prime} \,\right]_0 \bar{x} \bar{E}^{(\bar{x})}\,.
\end{equation}
Note that in case of the singular transformation \eqref{singular.case} a nonlinear equation ($M=1$) is transformed into a linear one ($\bar{M}=0$) but its structure is nevertheless preserved.

%%%%%%%%%%%%%%%%%%%%%%%%%%%%%%%%%%%%%%%%%%%%%%%%%%%%%%%%%%%%%%%%%%%%%%%%%%%%%%%%%%%%%%%%%%%%%%%%%%%%%%%%%%%%%%%%%%%%

\subsubsection{Transformation of the linearized equation}

The transformation of the quantity $\tilde{x}$, defined by \eqref{x.tilde}, reads 
\begin{equation}
\label{x.tilde.transformation}
\tilde{x} \equiv \pm \frac{x}{ \left| x \right|^{\frac{M}{2}} } = \pm \frac{ \bar{f}(\bar{\Phi}) - \left.\bar{f}(\bar{\Phi})\right|_0 }{ \left| \bar{f}(\bar{\Phi}) - \left.\bar{f}(\bar{\Phi})\right|_0 \right|^{\frac{M}{2}} } \equiv \pm \sqrt{ \left| \bar{Q}_3 \right| } \bar{\tilde{x}}
\end{equation}
where $\bar{Q}_3$ is the limiting value 
\begin{equation}
\bar{Q}_3 = \lim\limits_{\bar{\Phi} \to \bar{\Phi}_0} \left[ \frac{ \left(\bar{f}(\bar{\Phi}) - \left.\bar{f}(\bar{\Phi})\right|_0 \right)^{2-M} }{ \left(\bar{\Phi} - \bar{\Phi}_0 \right)^{2-\bar{M} } } \right] \,.
\end{equation}
The latter can be easily calculated by using the knowledge of \eqref{M.values} and the l'Hospital rule. The result reads
\begin{equation}
\label{Q.3.values}
\bar{Q}_3 = \left\lbrace  
\begin{array}{cl}
	\left. \left(\bar{f}^\prime\right)^2 \right|_0 \text{ or } \left. \bar{f}^\prime \right|_0 &\text{ for the regular case \eqref{regular.case}} \\
	\frac{1}{2}\left. \bar{f}^{\prime\prime} \right|_0 &\text{ for the singular case \eqref{singular.case}}
\end{array}
\right. \,.
\end{equation}
Therefore for both the regular \eqref{regular.case} and the singular \eqref{singular.case} case $\bar{Q}_3$ is nonvanishing and nondiverging. Therefore the order of the magnitude of $\tilde{x}$ is preserved under both transformations.

Combining the transformation rules of $M$ and $C_2$, given by \eqref{Mteisenemine} and \eqref{C.2.teisenemine} respectively, reveals 
\begin{equation}
\label{2.M.C.2.transformation}
(2-M)C_2 = \left.\mathrm{e}^{-2\bar{\gamma}}\right|_0 \left( \frac{2-\bar{Q}_2}{\bar{Q}_1} - \bar{M} \right) \bar{C}_2 = \left.\mathrm{e}^{-2\bar{\gamma}}\right|_0 \left(2-\bar{M}\right)\bar{C}_2 \,.
\end{equation}
Note that in the context of the invariant first order small source condition \eqref{nonhyperbolic.potential.condition.invariant} the constant $M$ effectively plays the role of $\frac{1}{2}$ which makes the difference between the invariant and noninvariant first order small conditions, given by \eqref{nonhyperbolic.potential.condition.invariant} and \eqref{condition.for.C.2}, respectively. Namely, due to the definition \eqref{condition.for.C.2} of $C_2$ and the GR regime conditions \eqref{general.relativity.conditions.for.FLRW.cosmology} (see also \eqref{potential.condition}) in the Friedmann cosmology
\begin{equation}
\label{intuitive.reasoning}
(2-M)C_2 = - \frac{\mathcal{A}}{2\ell^2}\Bigg\lbrace (2-M)\underset{\mathclap{=0 \text{ if }M=1}}{\underbrace{\frac{\mathcal{I}_2^{\prime\prime}}{\mathcal{F}}}} + (2-M)\underset{\mathclap{=0\text{ if }M=0}}{\underbrace{\left( \frac{1}{\mathcal{F}} \right)^\prime \mathcal{I}_2^{\prime}}} \Bigg\rbrace = - \frac{\mathcal{A}}{\ell^2}\left( \frac{\mathcal{I}_2^{\prime\prime}}{\mathcal{F}} + \frac{1}{2}\left( \frac{1}{\mathcal{F}} \right)^\prime \mathcal{I}_2^{\prime} \right) \,.
\end{equation}
Let us stress that neither $C_2$ nor $M$ is defined via invariants. However combining these two gives us an expression that only gains a finite multiplier under the local Weyl rescaling \eqref{conformal.transformation} and under the scalar field redefinition \eqref{field.redefinition}. As suggested by \eqref{intuitive.reasoning} the expression $(2-M)C_2$ is practically the one introduced by the invariant first order small source condition \eqref{nonhyperbolic.potential.condition.invariant}.

The results \eqref{C.1.teisenemine}, \eqref{x.tilde.transformation} and \eqref{2.M.C.2.transformation} impose that the expression in the curly brackets of \eqref{linear.perturbed.equation}, hence also the linearized equation transforms as follows
\begin{equation}
\ddot{\tilde{x}} + C_1^{\varepsilon} \dot{\tilde{x}} - \frac{2-M}{2}C_2\tilde{x} = \pm \sqrt{|Q_3|} \left.\mathrm{e}^{-2\bar{\gamma}}\right|_0 \left\lbrace \ddot{\bar{\tilde{x}}} + \bar{C}_1^{\varepsilon} \dot{\bar{\tilde{x}}} - \frac{2-\bar{M}}{2}\bar{C}_2 \bar{\tilde{x}} \right\rbrace \,.
\end{equation}

Let us study the transformation of $(2-M)C_2$ more in detail. Instead of considering the intermediate results \eqref{Mteisenemine} and \eqref{C.2.teisenemine} let us write the quantity $(2-M)C_2$ using the definitions for $C_2$, $M$ and invariants, given by \eqref{C.1.and.C.2}, \eqref{M} and \eqref{invariants} respectively, as follows
\begin{align}
\nonumber
(2-M)C_2 &= - \left.\frac{ \mathcal{A} }{ 2\ell^2 }\right|_0 \lim\limits_{x\to 0} \left\lbrace \left(2 - \left( \frac{1}{\left(\mathcal{I}_3^\prime\right)^2} \right)^\prime \left(\mathcal{I}_3^\prime\right)^2 x\right) \left( \frac{\frac{\mathcal{I}_2^\prime}{\left(\mathcal{I}_3^\prime\right)^2} }{x} \right) \right\rbrace = \\
\label{M.2.C.2.as.invariant}
&= - \left.\frac{ \mathcal{A} }{ \ell^2 }\right|_0 \left.\left\lbrace \frac{\mathcal{I}_2^{\prime\prime}\mathcal{I}_3^\prime}{ \left(\mathcal{I}_3^\prime\right)^3 } - \frac{ \mathcal{I}_2^\prime \mathcal{I}_3^{\prime\prime} }{ \left(\mathcal{I}_3^\prime\right)^3} \right\rbrace\right|_0 = - \left.\frac{ \mathcal{A} }{ \ell^2 }\right|_0 \left.\left\lbrace \frac{1}{\mathcal{I}_3^\prime}\left( \frac{\mathcal{I}_2^\prime}{ \mathcal{I}_3^\prime } \right)^\prime\right\rbrace\right|_0 \,.
\end{align}
Hence $(2-M)C_2$ indeed is the invariant first order small source condition \eqref{nonhyperbolic.potential.condition.invariant}. Such analogous procedure can be also carried through in case of the condition \eqref{matter.condition}. The previous results suggest that including the nonlinear term $\frac{M}{2}\frac{\dot{x}^2}{x}$ is an inevitable step.

\subsubsection{Transformation of the solutions}

A straightforward calculation reveals that due to the previously given transformation rules \eqref{C.1.teisenemine} and \eqref{2.M.C.2.transformation} the transformation of the eigenvalues \eqref{eigenvalue} read
\begin{equation}
\lambda^{\varepsilon}_\pm = \left. \mathrm{e}^{-\bar{\gamma}} \right|_0 \bar{\lambda}^{\varepsilon}_{\pm} \,.
\end{equation}
The result \eqref{M.2.C.2.as.invariant} allows to write the eigenvalues \eqref{eigenvalue} as \cite{JKSV:2}
\begin{equation}
\lambda^{\varepsilon}_{\pm} = \left[ \frac{\sqrt{\mathcal{A}}}{2\ell}\left( -\varepsilon \sqrt{3\mathcal{I}_2}  \pm \sqrt{ 3\mathcal{I}_2   -2   \frac{1}{\mathcal{I}_3^\prime}\left( \frac{\mathcal{I}_2^\prime}{ \mathcal{I}_3^\prime } \right)^\prime} \right) \right]_0 \,
\end{equation}
making the transformation properties obvious.

In order to obtain the transformation of the solution \eqref{solution.for.x}
\begin{equation}
\bar{x}(\bar{t}) = \pm \left( \bar{K}_1 \mathrm{e}^{\bar{\lambda}^{\bar{\varepsilon}}_{+} \bar{t}} + \bar{K}_2 \mathrm{e}^{ \bar{\lambda}^{\bar{\varepsilon}}_{-} \bar{t}} \right)^{\frac{2}{2-\bar{\rule[0.8ex]{0ex}{0ex}M}}}  \,
\end{equation}
in addition to the eigenvalues one must also consider the transformation \eqref{time.coordinate.transformation} of the cosmological time $t$. In \eqref{time.coordinate.transformation} only the transformation of the time coordinate differential is given. Here we are interested in the transformations calculated at the critical point. Hence in the lowest approximation level when considering the integral $\int \mathrm{e}^{\bar{\gamma}} \mathrm{d}\bar{t}$ we may assume the scalar field to be approximately constant. In other words $t = \left. \mathrm{e}^{\bar{\gamma}} \right|_0 \bar{t}$. Therefore
\begin{equation}
t \cdot \lambda^{\varepsilon}_\pm = \bar{t} \cdot \bar{\lambda}^{\varepsilon}_{\pm} \,.
\end{equation}
Hence the quantity $t\cdot\lambda^{\varepsilon}_{\pm}$, i.e.~the power of the exponent in the solution \eqref{solution.for.x}, gets transformed into itself.

Last but not least we have to consider the transformation of the power $\frac{2}{2 - M}$ of the solution \eqref{solution.for.x}. In the regular case \eqref{regular.case} $M = \bar{M}$ and the power does not change. However in the singular case \eqref{singular.case} $M = \frac{1}{2}\bar{M} + 1$ and hence 
\begin{equation}
\frac{2}{2 - M} = \frac{2}{2 - \frac{1}{2}\bar{M} - 1} = 2 \frac{2}{2-\bar{M}} \,.
\end{equation}
Therefore the power of the solution \eqref{solution.for.x} for $x$ is twice the one for $\bar{x}$. The latter is in perfect agreement with the transformation of the small perturbation in the singular case \eqref{singular.case}, i.e.\ \eqref{singular:x.etc.} where it was pointed out that $x \sim \bar{x}^2$. The difference of the power is a mathematical artefact due to the mapping between nonlinear and linear approximate equations, both covered by \eqref{First.order.equation}. One should keep in mind that in the generic parametrization the value of the scalar field $\Phi$ itself is not measurable and the physical meaning of the scalar field is not the same for different parametrizations. For example the JF BDBW parametrization \eqref{Jordan.frame} scalar field $\Psi$ encodes the nonminimal coupling $\Psi = \frac{1}{\mathcal{I}_1}$ while the EF canonical parametrization \eqref{Einstein.frame} scalar field $\varphi$ encodes the scalar field space volume \cite{JKSV:2}. 

However, leaving aside the power, the characteristic behaviour of a solution, i.e. convergence to GR regime or divergence from it, is determined by the eigenvalues \eqref{eigenvalue} and is therefore preserved under the local Weyl rescaling of the metric tensor \eqref{conformal.transformation} and under the scalar field redefinition \eqref{field.redefinition} even if the latter is singular.

%%%%%%%%%%%%%%%%%%%%%%%%%%%%%%%%%%%%%%%%%%%%%%%%%%%%%%%%%%%%%%%%%%%%%%%%%%%%%%%%%%%%%%%%%%%%%%%%%%%%%%%%%%%%%%%%%%%%%%%%%%%%%%%%%%%%%%%%%%%%%%%%%%%%%%%%%%%%%%%%%%%%%%%%%%%%%%%%%%%%%%%%%%%%%%%

\section{Summary}

We investigated first generation scalar-tensor theories of gravity characterized by four arbitrary coupling functions $\left\lbrace \mathcal{A}\,,\, \mathcal{B}\,,\, \mathcal{V}\,,\, \alpha\right\rbrace$ and invariant under the local rescaling of the metric and scalar field redefinition \eqref{conformal.transformation}-\eqref{field.redefinition}, \eqref{fl.fnide.teisendused}. Our main focus was upon the GR regime where the scalar field evolution has ceased and the remaining dynamical degrees of freedom are identical to those of GR. It is well known that in the GR regime the scalar field redefinition \eqref{field.redefinition} connecting JF BDBW \eqref{Jordan.frame} and EF canonical \eqref{Einstein.frame} parametrization is singular. As we pointed out in the introduction this singularity is physically meaningful and not due to an unfortunate choice of coupling functions. Therefore, for showing the equivalence of the parametrizations it is also important to study the transformation properties in the case of a singular scalar field redefinition.

In Section~\ref{general.theory} we started with general action functional \eqref{fl.moju} and derived the equations of motion for the metric tensor $g_{\mu\nu}$ \eqref{tensor.equation}, for the scalar field $\Phi$ \eqref{scalar.field.equation.without.R} and the matter continuity equation \eqref{EI.jaavus}. Specifying the FLRW line element gave the equations of motion \eqref{First.Friedmann.general.equation}-\eqref{cosmology.continuity.equation} in the Friedmann cosmology. By \eqref{transformation.of.tensor.equation}, \eqref{transformation.of.scalar.field.equation.without.R}, \eqref{transformation.of.continuity.equation} we showed how under the transformations these basic equations gain an overall multiplicative term containing the transformation functions of the metric rescaling and of the scalar field redefinition. To facilitate further discussion we also recalled the invariants \eqref{invariants} introduced in our earlier paper \cite{JKSV:2}. 

Section~\ref{general.relativity.regime} was concentrating on the GR regime, defined by \eqref{vanishing.nabla.Phi}, \eqref{potential.condition}-\eqref{matter.condition}. This definition is supplemented by assumptions \eqref{conditions.on.A.etc}-\eqref{conditions.on.A.prime.etc} that enforce the consistent notion of the GR regime and complementary restrictions \eqref{nonvanishing.F}-\eqref{non.vanishing.first.derivative.of.1.over.F}, \eqref{conditions.on.A.second.etc}-\eqref{nonvanishing.I.1.second.etc} necessary to make the corresponding critical point hyperbolic. To satisfy these conditions the allowed transformation functions fall into two cases, regular \eqref{regular.case} and singular \eqref{singular.case}. These results were used to show that the notion of the GR regime is invariant under the local Weyl rescaling and the scalar field redefinition. 

In Section~\ref{dynamical.system} we considered small perturbations of the scalar field \eqref{x} in the neighbourhood of the GR regime in the context of potential dominated Friedmann cosmology. It turned out that the perturbed equations \eqref{First.order.equation} in different parametrizations are in correspondence despite the fact that this equation itself might be nonlinear in one parametrization and linear in some other, related by a singular transformation \eqref{singular.case} giving relations \eqref{singular:x.etc.}. For instance the perturbed equation in JF BDBW parametrization in the case when $\omega$ diverges is nonlinear, while the corresponding perturbed equation in EF canonical parametrization is linear. These results complement our recent paper \cite{JKSV:2} where a slightly different approach was used. Last but not least we showed that the qualitative behaviour of the solutions, i.e.~whether the theory converges to general relativity regime or repels from it is independent of the parametrization.

To sum up, we demonstrated that if the general relativity  regime as a hyperbolic critical point is under consideration then there is an exact correspondence between different parametrizations even if the scalar field redefinition connecting these is singular. However in the latter case it is rather important to note that the order of magnitude of the small perturbation of the scalar field around some constant value changes under the singular scalar field redefinition as in \eqref{singular:x.etc.}. 

From a more general viewpoint we have developed a methodology which rather rigorously allows to check whether the imposed conditions are sufficient for establishing the equivalence of parametrizations. It would be interesting to study whether the correspondence is preserved if the conditions \eqref{condition.for.C.2}-\eqref{condition.for.C.3} leading to the hyperbolic critical point are loosened.

As an outlook it would be interesting as well to study the transformation properties in the context of second and third generation scalar-tensor theories \cite{Horndeski,disformal} while generalizing the local Weyl rescaling (conformal transformation) of the metric tensor to disformal transformation \cite{disformal:2}.
 
%%%%%%%%%%%%%%%%%%%%%%%%%%%%%%%%%%%%%%%%%%%%%%%%%%%%%%%%%%%%%%%%%%%%%%%%%%%%%%%%%%%%%%%%%%%%%%%%%%%%%%%%%%%%%%%%%%%%%%%%%%%%%%%%%%%%%%%%%%%%%%%%%%%%%%%%%%%%%%%%%%%%%%%%%%%%%%%%%%%%%%%%%%%%%%%%%%%%%%%%%%%%%%%%%%%%%%%%%%%%%%%%%%%%%%%%%%%%%%%%%%%%%%%%%%%%%%%%%%%%%%%%%%%%%%%%%%%%%%%%%%%%%%%%%%%%%%%%%%%%%%%%%%%%%%%%%%%%%%%%%%%%%%%%%%%%%%%%%%%%%%%%%%%%%%%%%%%%%%%%%%%%%%%%%%%%%%%%%%%%%%%%%%%%%%%%%%%%%%%%%%%%%%%%%%%%%%%%%%%%%%%%%%%%%%%%%%%

\bigskip \bigskip

{\bf Acknowledgments}

%\section{Acknowledgments}
%\ack sets the acknowledgments heading as an unnumbered section

This work was supported by the Estonian Science Foundation Grant No.~8837, by the Estonian Research Council Grant No.~IUT02-27 and by the European Union through the European Regional Development Fund (Project No. 3.2.0101.11-0029).

%i) Estonian Science Foundation, FundRef ID http://dx.doi.org/10.13039/501100001837 (Republic of Estonia)
%ii) Estonian Ministry for Education and Science Institutional Research, FundRef ID
%http://dx.doi.org/10.13039/501100002301 (Republic of Estonia)
%iii) European Union through the European Regional Development

%%%%%%%%%%%%%%%%%%%%%%%%%%%%%%%%%%5

\end{document}